\documentclass[
reprint,
amsmath,amssymb,
aip,
jap,
floatfix,
superscriptaddress
]{revtex4-1}
\usepackage{graphicx}					
\usepackage{dcolumn}					
\usepackage{bm}						
\usepackage{float}
\usepackage{gensymb}
\usepackage{multirow}
\usepackage{hyperref}
\usepackage{mathrsfs}
\usepackage{braket}
\usepackage[dvipsnames]{xcolor}

\newcommand{\im}{\mathrm{i}}

\newcommand{\Fig}[1]{Fig.~\ref{#1}}
\newcommand{\Eq}[1]{Eq.~\ref{#1}}
\newcommand{\Eqs}[2]{Eqs.~\ref{#1} and \ref{#2}}
\newcommand{\Eqsd}[2]{Eqs.~\ref{#1}-\ref{#2}}

\newcommand{\bvec}[1]{\bm{#1}}

\newcommand{\xhat}{\bvec{\hat{x}}}
\newcommand{\yhat}{\bvec{\hat{y}}}
\newcommand{\zhat}{\bvec{\hat{z}}}
\newcommand{\mhat}{\bvec{\hat{m}}}
\newcommand{\nhat}{\bvec{\hat{n}}}
\newcommand{\rhat}{\bvec{\hat{r}}}

\newcommand{\Ehat}{\bvec{\hat{E}}}

\newcommand{\kvec}{\bvec{k}}
\newcommand{\kp}{\kvec_{\bvec{||}}}

\newcommand{\phiR}{\phi_\text{r}}
\newcommand{\phiMP}{\phi_\text{mp}}

\newcommand{\UA}{\uparrow}
\newcommand{\DA}{\downarrow}
\newcommand{\UD}{\uparrow/\downarrow}
\newcommand{\uUD}{u_{\UD}}
\newcommand{\uU}{u_{\UA}}
\newcommand{\uD}{u_{\DA}}
\newcommand{\kzB}{k^*_z}

\newcommand{\ang}{\alpha}

\newcommand{\inc}{\text{I}}
\newcommand{\refl}{\text{R}}
\newcommand{\trans}{\text{T}}

\newcommand{\uangle}{\alpha}
\newcommand{\fsf}{\text{fs}}
\newcommand{\fcf}{\text{fc}}
\newcommand{\fs}{\fsf(\uangle)}
\newcommand{\fc}{\fcf(\uangle)}

\newcommand{\pu}{q}

\newcommand{\irl}{\text{i-r}}
\newcommand{\tl}{\text{t}}
\newcommand{\irtl}{\nu}

\newcommand{\SM}{S}
\newcommand{\SMir}{\SM_\irl}
\newcommand{\SMt}{\SM_\tl}
\newcommand{\SMirt}{\SM_\irtl}
\newcommand{\SMp}{\bar{\SM}}
\newcommand{\SMm}{\Delta\SM}

\newcommand{\fourvec}[1]{\mathfrak{#1}}

\begin{document}

\title{Interfacial spin-orbit torques}


\author{V. P. Amin}
\email{vpamin@iu.edu}
\affiliation{
Department of Chemistry \& Biochemistry, University of Maryland, College Park, MD 20742
}
\affiliation{
National Institute of Standards and Technology, Gaithersburg, Maryland 20899, USA
}
\author{P. M. Haney}
\email{paul.haney@nist.gov}
\affiliation{
National Institute of Standards and Technology, Gaithersburg, Maryland 20899, USA
}
\author{M. D. Stiles}
\email{mark.stiles@nist.gov}
\affiliation{
National Institute of Standards and Technology, Gaithersburg, Maryland 20899, USA
}

\date{\today}


\begin{abstract}

Spin-orbit torques offer a promising mechanism for electrically controlling magnetization dynamics in nanoscale heterostructures. While spin-orbit torques occur predominately at interfaces, the physical mechanisms underlying these torques can originate in both the bulk layers and at interfaces. Classifying spin-orbit torques based on the region that they originate in provides clues as to how to optimize the effect. While most bulk spin-orbit torque contributions are well studied, many of the interfacial contributions allowed by symmetry have yet to be fully explored theoretically and experimentally. To facilitate progress, we review interfacial spin-orbit torques from a semiclassical viewpoint and relate these contributions to recent experimental results. Within the same model, we show the relationship between different interface transport parameters. For charges and spins flowing perpendicular to the interface, interfacial spin-orbit coupling both modifies the mixing conductance of magnetoelectronic circuit theory and gives rise to spin memory loss. For in-plane electric fields, interfacial spin-orbit coupling gives rise to torques described by spin-orbit filtering, spin swapping and precession. In addition, these same interfacial processes generate spin currents that flow into the non-magnetic layer. For in-plane electric fields in trilayer structures, the spin currents generated at the interface between one ferromagnetic layer and the non-magnetic spacer layer can propagate through the non-magnetic layer to produce novel torques on the other ferromagnetic layer.
\end{abstract}

\maketitle

\section{Introduction}
\label{sec:introduction}
\begin{figure}[b!]
	\centering
	\vspace{0pt}	
	\includegraphics[width=1\linewidth,trim={0.7cm 0.0cm 0.8cm 0.0cm},clip]{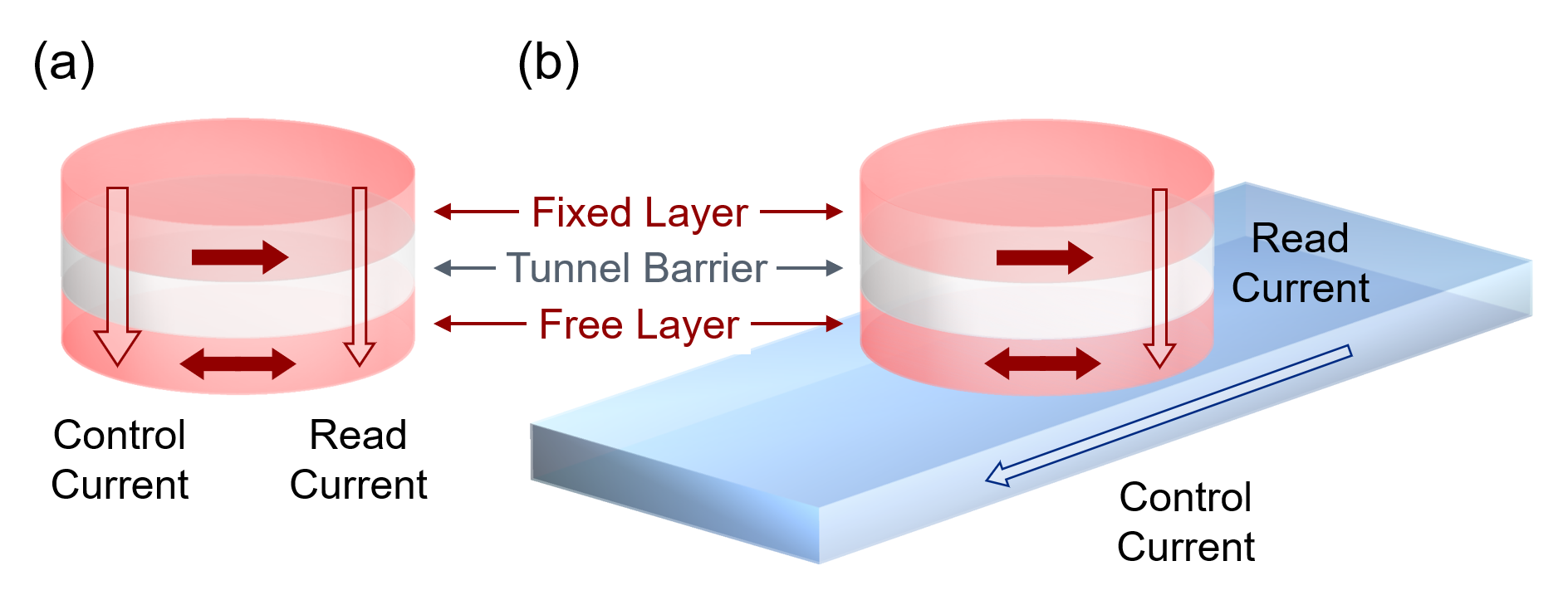}
	\caption{
    Magnetic tunnel junctions (dark red arrows represent magnetization direction). (a) Standard magnetic tunnel junction with fixed and free layers and the control current following the same path as the read current. (b) magnetic tunnel junction grown on heavy metal layer with separate read and control current paths.
	}
	\vspace{0pt}
	\label{fig:MTJs}
\end{figure}

Spintronic devices can augment modern integrated circuits with novel functionality, as exemplified by magnetoresistive random access memories. However, widespread adoption of additional spintronic devices depends on reducing the energy these devices require to control their magnetization dynamics via electrical currents.\cite{apalkov2016magnetoresistive,hanyu2016standby} Most commercial uses and many anticipated applications of spintronic devices are based on magnetic tunnel junctions because their large magnetoresistance\cite{butler2001spin,parkin2004giant,yuasa2004giant} makes it easy to measure their configuration. In most cases, the magnetization of one layer is fixed and the magnetization of the other layer is manipulated electrically. For manipulating the magnetization direction, all-electrical methods are preferred due to their compatibility with conventional electronic devices. In most devices, the control current typically flows across the tunnel junctions along the same path as the read current, see Fig.~\ref{fig:MTJs}(a). Such devices have challenging fabrication margins because the current that flows through the tunnel barrier must be much smaller than the current that can cause breakdown of the barrier. 

An alternative geometry was proposed about a decade ago in which the read currents also flow out-of-plane, but in which the control currents flow in-plane through a non-magnetic layer, usually a heavy metal, grown underneath the tunnel junction, see Fig.~\ref{fig:MTJs}(b). The torques in this geometry are called spin-orbit torques because spin-orbit coupling either in the interior of the layers or at the interfaces between them plays an essential role. By using these torques, such structures reduce the maximum current flow through the barrier and all but eliminate the problem of breakdown, while increasing the design space of possible devices.\cite{lee2016emerging} 

Optimizing the electrical control of magnetization could allow for a variety of new commercial applications of magnetic tunnel junctions. For applications in magnetic random access memory, the alternate geometry has a disadvantage compared to the original geometry because as a three terminal device it takes up more space on the chip.  On the other hand, there are indications that it switches faster. Due to this tradeoff between footprint and speed, the traditional and alternative geometries may be better suited for different applications, such as different levels of cache memory.\cite{lee2016emerging} Another domain of potential applications is in neuromorphic computing, where magnetic tunnel junctions can be used as local memory, superparamagnetic tunnel junctions, and spin torque nano-oscillators.\cite{grollier2016spintronic,grollier2020neuromorphic} Since one of the main driving forces for neuromorphic computing is reducing the energy consumption for different cognitive computing tasks, reducing the control current by optimizing spin-orbit torques becomes a key goal for spintronics-based approaches.

Here, we focus on controlling the free layer magnetization in a magnetic tunnel junction by passing current through an adjacent non-magnetic metal. We ignore the fixed layer and the tunnel barrier of the magnetic tunnel junction and focus on the magnetic free layer and the adjacent non-magnetic layer, referring to this pair as a bilayer structure. In addition, we consider a trilayer structure, in which an additional magnetic layer, not part of the magnetic tunnel junction, is added below the non-magnetic layer. This trilayer structure, sometimes called a spin valve, allows for non-zero torques on the free layer magnetization when symmetry requires that these torques be zero in bilayer structures.

The interfaces between layers play a fundamental role in spin-orbit torques.  They break inversion symmetry, as is necessary to generate a net torque on the magnetization. In addition, the reduced symmetry at the interface can enhance the role of spin-orbit coupling there, giving rise to interfacial coupling between the electric current and the spins. The goal of the paper is to provide understanding of the interfacial contributions to the spin-orbit torques in these bilayer and trilayer structures. Hopefully, this understanding will help lead to a reduction of the energy consumption for a variety of applications.

Spin-orbit torques have two classes of mechanisms, those due to spin-orbit coupling in the interior of the layers, called bulk mechanisms, and those due to spin-orbit coupling at the interfaces between layers, called interfacial mechanisms. The first reported observation of a spin-orbit torque was an observation of modified damping in a bilayer composed of a ferromagnet and a heavy metal.\cite{ando2008electric} The authors interpreted the mechanism as the heavy metal layer generating an out-of-plane spin current under the applied in-plane electric field from the spin Hall effect.\cite{SHETheoryDyakonovPerel,SHETheoryHirsch,SHETheoryZhang} That spin current exerts a spin transfer torque\cite{STTTheorySlonczewski2,STTTheoryBerger,stiles2006spin,STTTheoryRalph} upon flowing into the ferromagnetic layer. The mechanism was the motivation for an early experiment demonstrating the excitation of precessional dynamics through a spin-orbit torque.\cite{liu2012current} 

The prediction of an interfacial mechanism\cite{SOTTheoryManchon} for spin-orbit torques was based on the Rashba-Edelstein effect.\cite{REETheoryEdelstein} In this model, the two thin films are viewed as a two-dimensional electron gas. Electrons in this two-dimensional gas become spin-polarized under the applied in-plane electric field; these spin polarized electrons then exert torques on the magnetization of the ferromagnetic layer via the exchange interaction. For the first observation of switching due to spin-orbit torques,\cite{SOTExpMiron} the authors invoked this prediction to explain their results.  
In both the bulk and interfacial mechanisms, the applied in-plane electric field results in a torque on the magnetization, but the physical mechanism and the qualitative nature of the torque differ significantly. For a comprehensive review of theoretical and experimental progress on spin-orbit torques since then, see Ref.~\onlinecite{manchon2019current}. In the present review, we focus on a pedagogical description interfacial contributions to spin-orbit torques.


There has only been limited research addressing the role of interfacial spin-orbit coupling. Experimentally, it is difficult to distinguish between bulk and interfacial mechanisms of spin-orbit torques because there is no difference in the symmetry of the resulting torques. One can only hope to differentiate them through indirect measurements like thickness dependence or material variations. Unfortunately, doing so through such measurements requires that other properties of the sample do not change as the thickness or materials are varied, which is almost never the case. In addition, as we discuss below, the importance of multiple length scales can make it difficult to interpret the experiments.

First principles calculations of spin-orbit torques\cite{SOTTTheoryHaney2,SOTTheoryFreimuth,SOTTheoryFreimuth2,SOTTheoryFreimuth3,SOTTheoryGeranton,iSOCAminiGSC,mahfouzi2018first,belashchenko2019first,belashchenko2020interfacial} naturally include the processes that contribute to both bulk and interfacial mechanisms. Unfortunately, they are not at a state where they can definitively identify the origin of the torques. These calculations are numerically intensive, so that few systematic thickness and material studies have been done.\cite{SOTTheoryFreimuth,SOTTheoryFreimuth2,belashchenko2019first,belashchenko2020interfacial} Of those, some but not all suggest interfacial contributions. Most experimental systems are quite disordered and disorder is difficult to treat in first principles calculations. Furthermore, the types of disorder that can be treated do not necessarily reflect the relevant experimental systems. Including onsite disorder\cite{SHETheoryWang,belashchenko2019first} allows calculated systems to have the high resistivities measured experimentally, but it is unclear how effectively such calculations capture the role of structural disorder, including amorphous structures, polycrystallinity, and grain boundaries that may be important in these systems.

In this paper we adopt a semiclassical approach,\cite{SOTTheoryHaney,iSOCAminFormalism,iSOCAminPhenomenology} which despite of some disadvantages compared to a first principles approach offers significant advantages for pedagogy. Semiclassical calculations are based on assumptions that are seldom justified in these systems. They assume that system sizes and scattering lengths are much larger than the electron wavelengths. However, layer thicknesses in experimentally-relevant systems tend to approach that length scale. Semiclassical approximations leave out quantum interference effects, though these effects have not been observed experimentally in connection with spin-orbit torque. An additional drawback of semiclassical approximations is an explosion of parameters that are not all constrained by experiment. On the other hand, semiclassical calculations are be easier to interpret than first principles calculations and offer a clear separation between bulk and interface effects. 

The most common semiclassical approach is the drift-diffusion approximation in which the system is described in terms of densities and currents. While this approach is often considered the most natural way to describe experimental results, there are at least two reasons to use a description based on the Boltzmann equation. The first is that since the early theories of current-in-plane giant magnetoresistance\cite{CamleyBarnas} it is known that the appropriate length scale for in-plane transport is the mean free path rather than the spin diffusion length. Variations on the length scale of the mean free path are captured by the Boltzmann equation but not drift diffusion approaches. More importantly, since spin-orbit coupling couples the electron spin to its motion, a wave-vector-dependent approach is needed to capture its effects.  In this work, we start with simple model described by the Boltzmann equation and show how that model connects to the parameters that might enter a description based on the drift-diffusion equation.  

Given the experimental and theoretical difficulties in distinguishing bulk and interfacial mechanisms for spin-orbit torques, what is the rationale for studying interfacial mechanisms?  The main reason is to develop a clear picture of what system properties lead to optimal behavior.  For example, an analysis of interfacial spin-orbit torques could help determine whether to minimize or maximize interfacial spin-orbit coupling or to minimize or maximize the bulk spin diffusion length. Another important reason to study interfacial mechanisms lies in recent experiments on trilayer structures driven by in-plane currents. In these systems, spin currents generated at the interfaces and/or the ferromagnetic layers enable additional functionality compared to bilayers, such as field-free switching of perpendicularly-magnetized layers.\cite{iSOCBaekAmin} Determining exactly what drives magnetization dynamics in these systems will offer new insights into the nature of spin-orbit torque. Both theory\cite{AHEAMRTaniguchi,iSOCAminiGSC,FSHEAmin} and experiment\cite{SOTExpHumphries,iSOCBaekAmin,iSOCHibino} suggest that bulk and interfacial mechanisms could play a role in these systems, but here the bulk mechanisms originate in the ferromagnetic layers rather than a heavy metal. Thus, disentangling bulk and interfacial contributions remains an important challenge, even as new device geometries are explored.

The goal of this paper is to provide a pedagogical explanation of interfacial contributions to spin-orbit torque, using a semiclassical approach.  In Section~\ref{sec:background}, we give background for subsequent discussions.  This background includes a discussion of the flow of angular momentum between reservoirs (Sec.~\ref{sec:angularMomentum}), the role that interfaces play in perpendicular transport (Sec.~\ref{sec:Perpendicular}) and in-plane transport both for bilayers (Sec.~\ref{sec:InPlane}) and trilayers (Sec.~\ref{sec:InPlaneTrilayers}), the distinctions between extrinsic and intrinsic mechanisms (Sec.~\ref{sec:extint}), the angular dependence of the torques that are allowed by symmetry (Sec.~\ref{sec:symmetry}), and complications associated with distinguishing bulk and interface contributions from the thickness dependence (Sec.~\ref{sec:lengthScales}). With that background, in Sec.~\ref{sec:mechanisms}, we use a highly simplified model to describe the different mechanisms that can generate interfacial contributions to spin-orbit torques.

\section{Background}
\label{sec:background}

\subsection{Angular Momentum}
\label{sec:angularMomentum}

\begin{figure}[b!]
	\centering
	\vspace{0pt}	
	\includegraphics[width=1\linewidth,trim={0.0cm 0.0cm 0.0cm 0.0cm},clip]{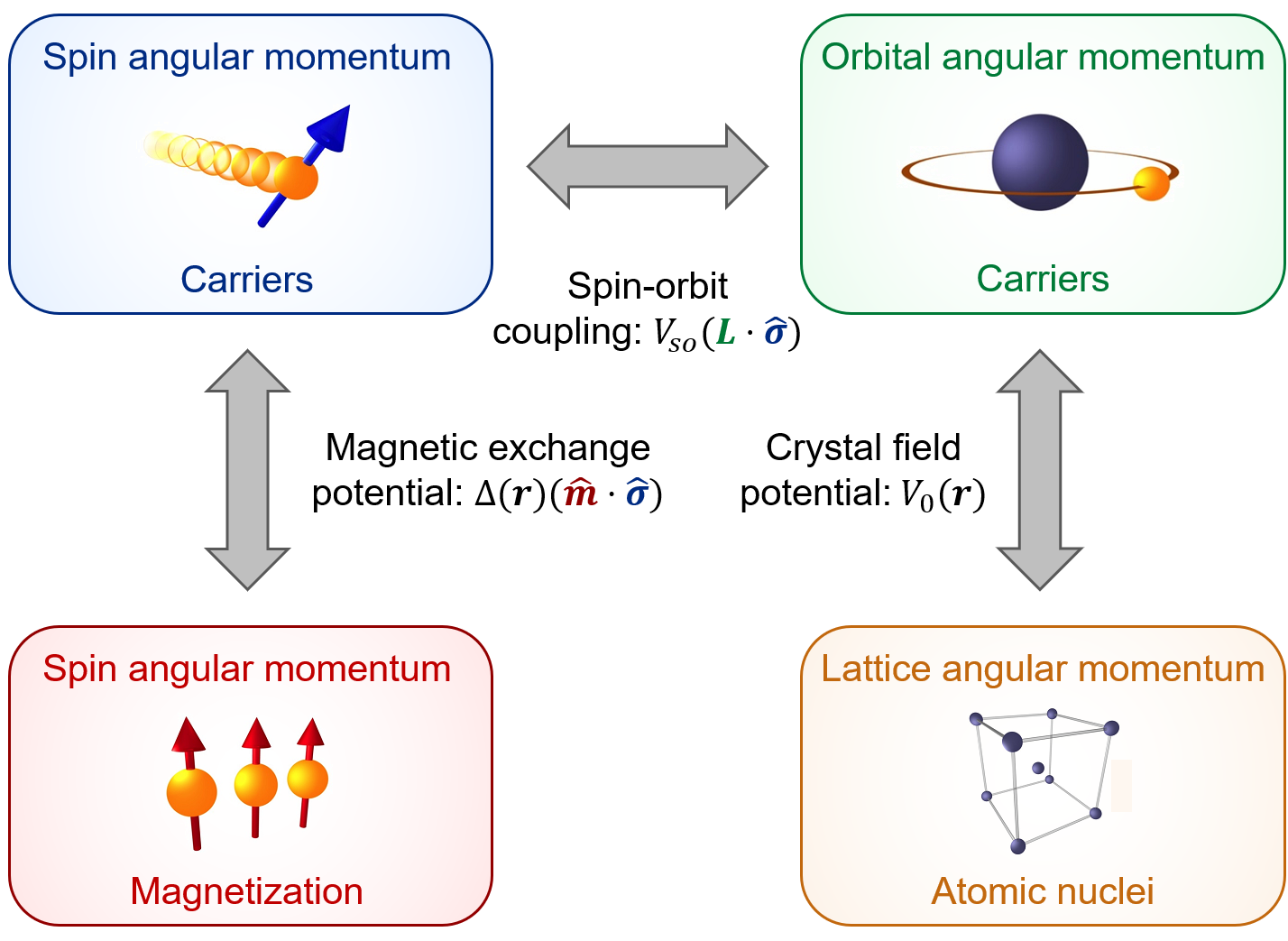}
	\caption{
    Schematic of different angular momentum reservoirs and the interactions coupling them. In ferromagnetic metals, the net magnetization is the sum of the magnetic moments of electrons carrying both orbital and spin angular momentum, with the latter dominating in transition metal ferromagnets. The magnetic exchange potential couples the spin angular momentum of the magnetization to the spin angular momentum of the carriers. The spin-orbit interaction couples the spin angular momentum of the carriers to their orbital angular momentum. The crystal field potential couples the orbital angular momentum of carriers to the angular momentum of the atomic lattice. Spin-orbit torques arise when an applied electric field promotes angular momentum transfer from the atomic lattice to the magnetization using carriers as mediators for the transfer.
	}
	\vspace{0pt}
	\label{Lreservoirs}
\end{figure}

Tracking the flow of angular momentum in the system provides a useful framework for understanding spin-orbit torques.  The total angular momentum of the system includes contributions from the ions comprising the lattice and the electrons, which possess an orbital angular momentum and an intrinsic angular momentum derived from their spin degree of freedom.  It is useful to further partition the electrons' spin angular momentum into a component from the magnetic order parameter ({\it e.g.}, the electrons comprising the magnetization condensate) and a component from non-equilibrium states participating in transport. Each of these components represent a reservoir of angular momentum, and our interest is in tracking the flow of angular momentum from these reservoirs to the magnetization upon the application of an electric field as shown in Fig.~\ref{Lreservoirs}.

The transfer of angular momentum between reservoirs is mediated by interactions, which are described in the Hamiltonian for the electrons:
\begin{eqnarray}
H = \frac{\hbar^2 \nabla^2}{2 m_e} +  V_0(\bm{r})+ \Delta(\bm{r}) \left( \bm{\hat m} \cdot \bm{\hat{\sigma}}\right)  + V_{so} \left(\bm{L} \cdot \bm{{\hat {\sigma}}}\right) . \label{eq:Hgeneral}
\end{eqnarray}
The first term is the kinetic energy. The second term $V_0(\bm{r})$ is the crystal field potential, which breaks rotational symmetry for the electrons, so that the angular momentum of electrons is not conserved.  This term enables the flow of angular momentum from the electronic system to the lattice. Note that the Hamiltonian for the whole system, including that of the lattice, is rotationally invariant, so that total angular momentum is conserved. The third term is the exchange interaction between electron spin $\bm{\sigma}$ and the magnetization, which is oriented in the $\bm{\hat m}$ direction.  Its magnitude $\Delta(\bm{r})$ is position dependent, and can be determined self-consistently in a mean-field theory approach, or taken as a constant in simpler models, such as the Stoner model. The fourth term is the spin-orbit coupling, where we only include contributions from the onsite, atomic-like form $\bm{L}\cdot \bm{\sigma}$, and parameterize its strength with $\alpha$.  This is the dominant source of spin-orbit coupling in most materials, owing to the rapid orbital motion (compared to linear motion) of electrons and the strong electric fields near the nucleus.

The degrees of freedom in Eq. \ref{eq:Hgeneral} represent the different reservoirs of angular momentum, while the coupling between degrees of freedom mediate the transfer of angular momentum between reservoirs, as shown schematically in  Fig.~\ref{Lreservoirs}.  Spin transfer torque, which we discuss in the following section, is a transfer of angular momentum between the magnetization and the electron spin of current-carrying electrons.  In systems with strong spin-orbit coupling, the magnetization is also coupled to the orbital angular momentum of the electrons and to the lattice, opening up a wider array of mechanisms for exerting torques on the magnetization. This framework of tracking angular momentum flow is quite general and described in more details in Refs.~\onlinecite{haney2010current,go2020first}.  

\subsection{Perpendicular Transport}
\label{sec:Perpendicular}

\begin{figure}
	\centering
	\vspace{0pt}	
	\includegraphics[width=1\linewidth,trim={0.0cm 0.0cm 0.0cm 0.0cm},clip]{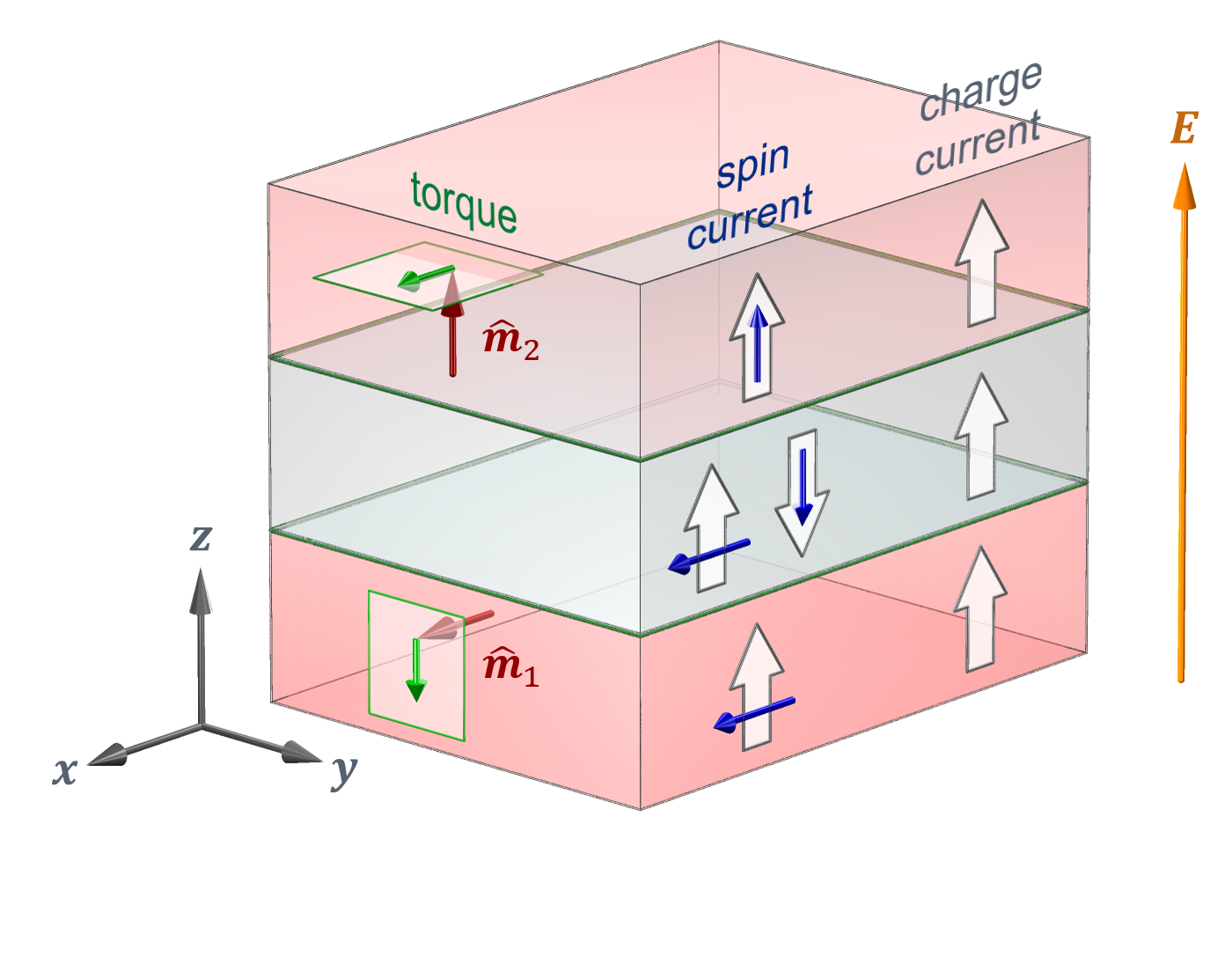}
	\caption{Magnetic trilayer and perpendicular transport. The top and bottom ferromagnetic layers are separated by a non-magnetic layer. In each layer, the charge current flows along the electric field, where flow directions are given as block arrows. In each ferromagnetic layer, the spin current flows along the charge current with spins aligned with the magnetization (red arrows). For spin currents, block arrows indicate electron flow direction and blue arrows indicate spin direction. Equivalently, block arrows could also indicate charge flow direction with blue arrows indicating magnetic moment. In the non-magnetic layer, the spins in the spin current are a combination of spins aligned with the lower layer magnetization and anti-aligned with the upper layer magnetization (here given by $\hat{\bm{x}}-\hat{\bm{z}}$). In the absence of spin-orbit coupling, the spin current with spin direction longitudinal to the magnetization is conserved across the interfaces. Note that spin currents are unchanged by flipping both the flow and spin directions. The discontinuity in the spin current at the interfaces, given by the spin direction transverse to the magnetization, is the spin transfer torque, indicated for the top and bottom layers by the green arrows. Typically, one layer will be able to respond to the torques and the other layer will be essentially fixed though one of several mechanisms.
	}
	\vspace{0pt}
	\label{TrilayerSTT}
\end{figure}

The study of perpendicular transport in magnetic multilayers (see Fig.~\ref{TrilayerSTT}) began with measurements of the current-perpendicular-to-the-plane giant magnetoresistance.\cite{pratt1991perpendicular,bass1999current} Following that, two intertwined phenomena dominated the field, spin transfer torques\cite{STTTheorySlonczewski2,STTTheoryBerger,tsoi1998excitation,myers1999current,katine2000current} and tunneling magnetoresistance.\cite{moodera1995large,butler2001spin,parkin2004giant,yuasa2004giant} In all of these, interfaces play a crucial role. For giant magnetoresistance, spin-dependent scattering at the interface leads to a spin-dependent interface resistance,\cite{bass1999current,schep1997interface,stiles2000calculation,} which can dominate the resistance for thin enough layers. This same spin-dependent scattering leads to a spin-transfer torque.

In magnetic multilayers or tunnel junctions, spin transfer torques are the torques on the magnetizations exerted by the spins of non-equilibrium, current-carrying electrons for currents flowing perpendicular to the plane of the layers.  These torques are generically present when an electric current is applied to a system where the magnetization is oriented differently in different layers. When electrons with spin-polarization aligned with one magnetic layer interact with a subsequent magnetic layer, two processes contribute to the torque.\cite{Stiles:2002a,STTTheoryRalph} The first is that the electron spins precess around the magnetization at the interface and exert a reaction torque on the magnetization.  The second is that the spin current that propagates into the ferromagnetic layer rapidly dephases and becomes aligned with the magnetization. These processes are discussed in more detail in Sec.~\ref{sec:mechanisms}.

The physics of spin transfer torque is most easily understood in the limit where spin-orbit coupling is small compared to the magnetic exchange energy.  In this case, an equation of continuity for total spin (magnetization plus conduction electron spin) relates the torque on a volume of magnetization to the net flux of transverse spin current into the volume. Magnetoelectronic circuit theory\cite{MCTBrataas,MCTBrataas2} provides a description of perpendicular transport when interfacial spin-orbit coupling is weak.

The first indication of the importance of interfacial spin-orbit coupling was the determination that some current-perpendicular-to-the-plane giant magnetoresistance measurements could not be adequately fit unless they allowed for finite spin relaxation at  interfaces\cite{galinon2005pd,bass2007spin,SPExpSanchez} rather than simply spread out through the layers.  Recent first-principles calculations\cite{dolui2017spin,gupta2020disorder} support this phenomenology. This is most dramatically illustrated by spin memory loss at interfaces between to normal metals.\cite{SMLExpKurt,SMLExpNguyen,dolui2017spin}  Since inversion symmetry is broken at such interfaces, special forms of spin-orbit coupling are allowed.  These cause wave-vector-dependent precession in the spin-orbit effective field and a reduction of spin current crossing the interface.

\subsection{In-plane Transport in Bilayers}
\label{sec:InPlane}

For bilayer systems composed of nonmagnetic and ferromagnetic layers, in-plane transport leads to torques on the magnetization from several distinct sources.  Although the bilayer geometry is simpler than that of trilayers, the materials are chosen to utilize spin-orbit coupling for generating torques.  This enlarges the set of reservoirs and interactions which contribute to the torque, so that identifying the different sources of torque is a more difficult task.  In this section we review the mechanisms of spin-orbit torque in this geometry. We first briefly describe the spin Hall and orbital Hall contributions, which arise from transporting angular momentum from the nonmagnetic layer to the ferromagnet.  We then discuss the recently discovered anomalous torque, and conclude with a longer discussion on interfacial torques.

\begin{figure}
	\centering
	\vspace{0pt}	
	\includegraphics[width=1\linewidth,trim={0.0cm 0.0cm 0.0cm 0.0cm},clip]{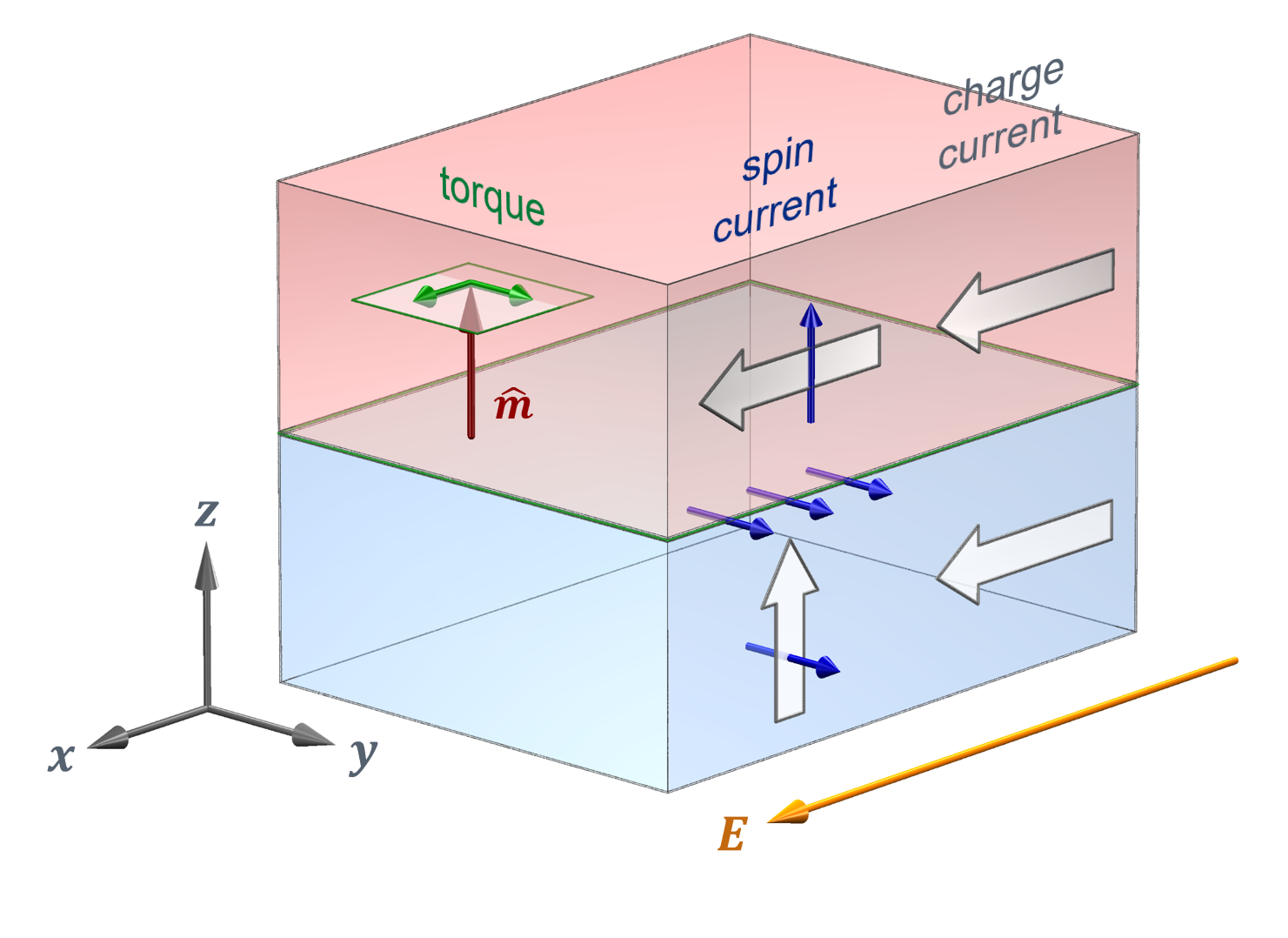}
	\caption{Magnetic bilayer and in-plane transport. In both the top ferromagnetic layer and bottom non-magnetic layer, charge currents flow along the electric field, where flow directions are given as block arrows. In the ferromagnetic layer, the spin current flows along the charge current with spins aligned with the magnetization (red arrow), where for spin currents block arrows give flow direction and blue arrows give spin direction. Green arrows indicate the two components of the torque on the magnetization. In the bottom nonmagnetic layer, the spin Hall effect generates a spin current with flow along $\hat{\bm{z}}$ and spin direction along $\hat{\bm{y}}$. In the absence of spin-orbit coupling at the interface, the discontinuity of the spin Hall current across the interface gives the interfacial contribution to spin-orbit torque on the magnetization. However, with nonvanishing interfacial spin-orbit coupling, the Rashba-Edelstein effect generates a spin accumulation at the interface that exerts an exchange torque on the magnetization. As will be discussed throughout this review article, additional torques arise from spin-orbit scattering at the interface (not shown here), possibly contributing to torques measured in experiments.
	}
	\vspace{0pt}
	\label{BilayerSOT}
\end{figure}

The spin Hall effect plus spin transfer torque mechanism was proposed to explain one of the early experiments on spin-orbit torques.\cite{liu2012current} For many of the systems studied to date, this mechanism is considered to provide the primary contribution to the dampinglike torque.  It is based on the spin Hall effect in the nonmagnetic layer, which results in a spin current which flows in all directions perpendicular to the electric field, with the spin directions perpendicular to both the spin flow and electric field directions. This effect was first predicted by D'yakonov and Perel\cite{SHETheoryDyakonovPerel} using a semiclassical approach and later explained using several other mechanisms,\cite{SHETheoryHirsch,SHETheoryZhang,SHETheoryMurakami,SHETheorySinova} eventually resulting in a mostly unified picture.\cite{SHEReviewSinova}  We refer interested readers to more in-depth reviews on the spin Hall effect.\cite{SHEReviewSinova,SHEReviewMurakami,SHEReviewHankiewicz,SHEReviewWunderlich,hoffmann2013spin} The spin current generated in the nonmagnetic layer is injected into the ferromagnet.  If the spin-orbit coupling at the interface and in the ferromagnet is much smaller than the exchange splitting, then the torque on the magnetization equals the incoming spin current due to the spin transfer torque mechanism.

The orbital Hall effect plus spin transfer torque is a more recently proposed mechanism of spin-orbit torque. In this case, the applied electric field induces orbital angular momentum flow in the nonmagnet, with similar symmetry properties to the spin Hall effect: the flow direction is perpendicular to the applied electric field, and the angular momentum direction is perpendicular to the field and flow directions.\cite{bernevig2005orbitronics,tanaka2008intrinsic,kontani2009giant,go2018intrinsic,jo2018gigantic} This orbital angular momentum is injected into the adjacent ferromagnet, where spin-orbit coupling in the ferromagnet transduces the orbital current to a spin accumulation, which exerts a torque on the magnetization.\cite{go2020orbital,go2020first} Experimentally distinguishing orbital Hall from spin Hall contributions is challenging, and is discussed in Ref.~\onlinecite{go2020first}. Note that orbital angular momentum also plays a crucial role in the spin Hall effect. The electric field does not couple directly to the electrons' spins but rather couples to their orbital moments. These in turn couple to the spins through spin-orbit coupling.

The anomalous torque is an effect in which the application of an electric field to a {\it single} ferromagnetic layer leads to torques in the ferromagnet where inversion symmetry is broken, for example, at interfaces.  Recent experimental work has confirmed that single layer ferromagnets experience spin torques at their layer boundaries under applied electric fields.\cite{ASOTWang,safranski2019spin} These \emph{anomalous} torques may arise from spin currents generated in the bulk with spin direction transverse to the magnetization. Theoretical studies\cite{FSHEAmin} show that such spin currents, which are allowed by symmetry and not subject to dephasing, are comparable in strength to spin Hall currents in Pt. When these spin currents flow to the layer boundaries, where inversion symmetry is broken, they can exert spin transfer torques. The same effect was studied in a different context in Ref.~\onlinecite{kurebayashi2014antidamping}, which proposed dampinglike spin-orbit torques in ferromagnets with broken bulk inversion symmetry, and experimentally observed these torques in a strained ferromagnetic semiconductor, GaMnAs. 

\begin{figure*}[t!]
	\centering
	\vspace{0pt}	
	\includegraphics[width=0.95\linewidth,trim={0.0cm 0.5cm 0.0cm 0.3cm},clip]{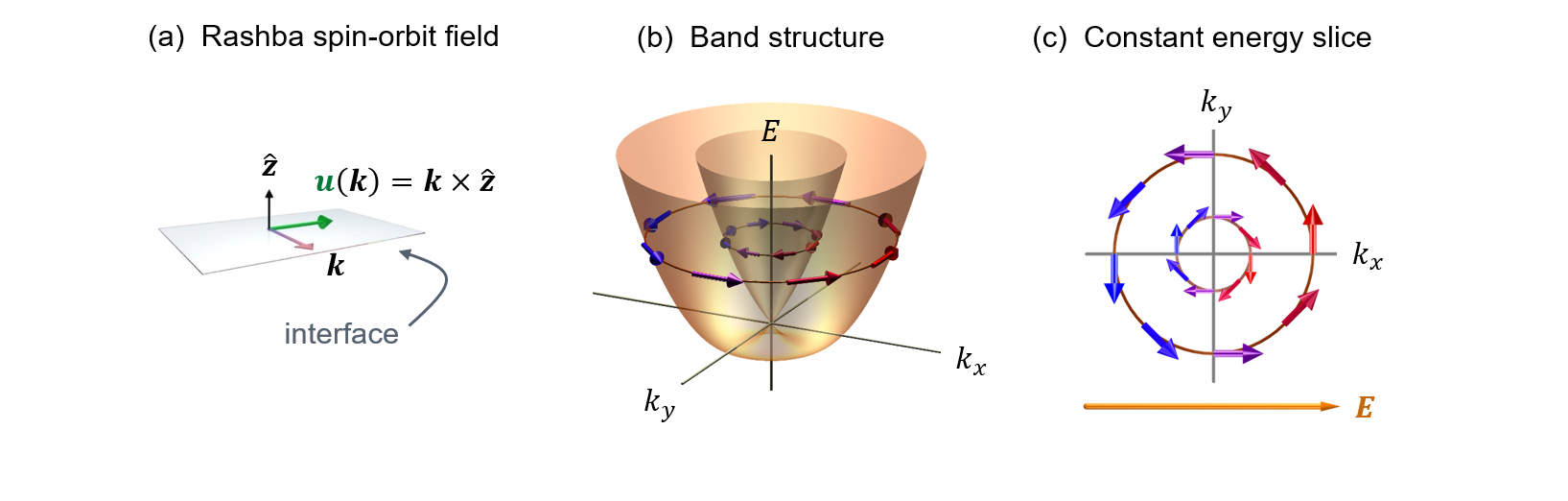}
	\caption{
	Real and reciprocal space depictions of the Rashba-Edelstein effect. Carriers are restricted to an idealized two-dimensional interface. (a) The carriers feel an effective magnetic field along $\bvec{u}(\bvec{k}) = \bvec{k} \times \zhat$ (green arrow) due to spin-orbit coupling. (b) The band structure obtained from the Rashba Hamiltonian in \Eq{eq:Hrashba} for vanishing exchange interaction ($\Delta = 0$). The spin expectation values (arrows) are shown at the Fermi energy $E_F$, where $E_F > E(\bvec{k} = \bvec{0})$. (c) The Fermi surface forms two circular sheets distinguished by their spin expectation values being parallel (outer circle) or antiparallel (inner circle) to $\bvec{u}(\bvec{k})$. An electric field biases carrier occupations, where blue arrows indicate increased occupation and red decreased, leading to a net spin polarization along $\bvec{E}\times\zhat$.
	}
	\vspace{0pt}
	\label{RashbaEdelstein}
\end{figure*}

\begin{figure}[t!]
	\centering
	\vspace{0pt}	
	\includegraphics[width=1.0\linewidth,trim={0.0cm 0.1cm 0.0cm 0.5cm},clip]{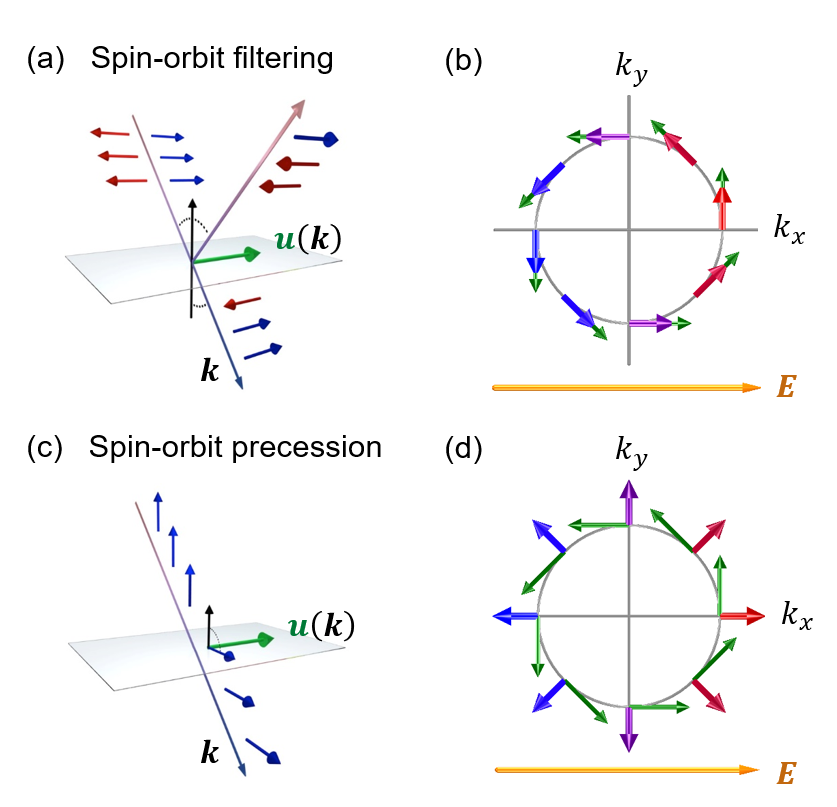}
	\caption{
	Real and reciprocal space depictions of the role of interfacial spin-orbit coupling.  Green arrows indicate the effective magnetic field along $\bvec{u}(\bvec{k}) = \bvec{k} \times \zhat$ due to spin-orbit coupling. In panels (a) and (c), red and blue arrows indicate spin directions. In panels (b) and (d), blue arrows indicate increased occupation due to the in-plane electric field and red decreased.  (a) Unpolarized carriers scattering from the interfacial spin-orbit field become spin-polarized (spin-orbit filtering) because the field creates a spin-dependent potential barrier. (b) Spin polarization after transmission through the interface for unpolarized incoming carriers on a circular slice (constant $k_z$) of one sheet of the Fermi surface.  The non-equilibrium occupation due to the electric field leads to a net flow of transmitted electrons along $\zhat$ with a net spin polarization along $\zhat \times \bvec{E}$. (c) In ferromagnetic layers, carriers are spin-polarized along the magnetization. Scattering from the interface, these spins precess around $\bvec{u}(\bvec{k})$ (spin-orbit precession). (d) In-plane spin polarization after transmission through the interface for incoming carriers polarized along $\zhat$ on a circular slice (constant $k_z$) of one sheet of the Fermi surface. The non-equilibrium occupation due to the electric field leads to the transmitted spins carrying a net spin flow along $\zhat$ with net spin polarization along $\mhat \times (\zhat \times \bvec{E})$.
	}
	\vspace{0pt}
	\label{From2Dto3D}
\end{figure}

In addition to these spin-orbit-torque mechanisms in which angular momentum is supplied from the interior of the layers, there are also contributions in which the angular momentum is supplied at the interfaces between layers. These interfacial contributions to the spin-orbit torque are the focus of this review.  In the initial model for spin-orbit torque\cite{SOTTheoryManchon}, the interface plays a direct role in generating a magnetic torque through interfacial spin-orbit coupling.  It's useful to study interfacial contributions by examining the Rashba model, which provides a minimal description of spin-orbit coupling in systems with broken structural inversion symmetry.  For broken inversion symmetry along the $\bm{z}$-direction, the Rashba interaction is given by $\bm{\sigma}\cdot\left(\bm{k}\times\bm{z}\right)$. For nonmagnetic systems, the Rashba interaction lifts the spin-degeneracy of states with nonzero Bloch wave vector $\bm{k}$. Electron states are still doubly degenerate (Kramer's doublet) but now the two degenerate states exist at $\bm{k}$ and $-\bm{k}$, with time reversal symmetry ensuring $\bm{s}(\bm{k})=-\bm{s}(-\bm{k})$. This degeneracy implies that the net spin density of the Rashba model without ferromagnetism or a magnetic field vanishes in equilibrium. However, under an applied electric field, the nonequilibrium occupation of carriers with wavevectors $\pm \bvec{k}$ differs in general, so a net nonequilibrium spin density (or spin accumulation) forms at the interface, as shown in Fig.~\ref{From2Dto3D}(b).

In ferromagnetic systems that lack structural inversion symmetry, a spin accumulation still forms in response to an in-plane electric field and this spin accumulation exerts an exchange torque on the magnetization. For broken inversion symmetry along the $\bm{\hat {z}}$-direction, the minimal Hamiltonian is
\begin{eqnarray}
H = \frac{\hbar^2 \nabla^2}{2 m_e} + V_0 + \Delta \left( \bm{\hat m} \cdot \bm{\hat{\sigma}}\right) + \alpha_R\bm{\sigma}\cdot\left(\bm{k} \times \bm{z}\right) .
\label{eq:Hrashba}
\end{eqnarray}
This Hamiltonian differs from that in Eq.~\ref{eq:Hgeneral} in that the crystal field potential and the atomic spin-orbit coupling have been replaced by the Rashba form of spin-orbit coupling, which is wave-vector dependent.  This transformation is based on the assumption that the wave vector perpendicular to the symmetry-breaking direction $\bm{\hat {z}} $ is a good quantum number and that all of the effects of the crystal-field potential can be absorbed into $V_0$, $\alpha_R$, and possibly an effective mass (see Ref.~\onlinecite{RasbhaTheoryPark}). In a multilayer, these parameters vary from layer to layer and $\alpha_R$ becomes large at interfaces where symmetry breaking is strongest.

Several studies\cite{SOTTTheoryHaney2,grytsyuk2016k} have addressed the relevance of simplified models, like the Rashba model, using density functional theory to describe bilayers in a slab geometry, in which several atomic layers are included away from the interfaces.  These calculations\cite{SOTTTheoryHaney2,grytsyuk2016k} show that the Rashba spin-orbit interaction, as measured by the misalignment between the spin and the local exchange field, is highly localized and dominant on the interfacial atoms.  These results provide a motivation for the models of interfacial Rashba spin-orbit coupling described below in Sec.~\ref{sec:mechanisms}.

While Rashba spin-orbit coupling is highly localized on the interfacial atoms, electronic transport is not confined to the interface, as assumed in the typical description of the Rashba-Edelstein effect. Although experiments and first principles calculations have confirmed that localized interface electronic states form at various material interfaces\cite{lashell1996spin,nicolay2001spin,cercellier2004spin,bendounan2011evolution}, the remaining electronic states in the bilayer are not confined to the interface plane. Therefore, it is useful to extend the Rashba-Edelstein model from two dimensions to three dimensions by treating the spin-dependent scattering of electrons off a localized interfacial potential\cite{SOTTheoryHaney,iSOCAminFormalism,iSOCAminPhenomenology}. Section~\ref{sec:mechanisms}  discusses such a calculation in detail.

To motivate the more complete discussion in Sec.~\ref{sec:mechanisms}, we start with a simple extension of the Rashba model from two to three dimensions\cite{iSOCBaekAmin,iSOCAminiGSC}. In this model, we omit an interfacial exchange interaction and assume the crystal field potential and Rashba potential are localized at the interface ($z = 0$):
\begin{eqnarray}
H = \frac{\hbar^2 \nabla^2}{2 m_e} + t \delta(z) \big{[}
V_0 + \alpha_R\bm{\sigma}\cdot\left(\bm{k} \times \bm{z}\right) 
\big{]}
\label{eq:HrashbaInt}
\end{eqnarray}
Here $z$ is the out-of-plane direction and $t$ is the relevant interfacial length scale. In what follows, we discuss what happens when free electrons from the bulk layers scatter off the interface, modeled by the delta function potential given in \Eq{eq:HrashbaInt}. Even under in-plane electric fields, carrier motion is largely isotropic, so the electron distribution functions within an average elastic scattering length (mean free path) of the interface are modified by interfacial scattering despite the formation of net in-plane currents in the bulk layers. In response to in-plane electric fields, the interfacial scattering leads to \emph{out-of-plane} spin currents that can exert spin torques on the ferromagnetic layer, as depicted in Fig.~\ref{From2Dto3D}(c)-(f).

Unpolarized free electrons from the bulk layers become spin polarized (for nonvanishing $V_0$ and $\alpha_R$) after scattering off the interface. This filtering effect occurs because the Rashba potential acts as a spin- and momentum-dependent potential barrier. In particular, the Rashba potential preferentially reflects or transmits electrons based on their spin, so an unpolarized stream of electrons becomes spin-polarized after scattering, as seen in Fig.~\ref{From2Dto3D}(c). However, in equilibrium, the net spin current vanishes after summing over all $\bvec{k}$-states, much like the vanishing equilibrium spin density under the conventional Rashba-Edelstein effect. However, the Rashba potential also depends on the momentum of incident electrons. In the presence of an in-plane, applied electric field $\bvec{E}$, the occupation of carriers becomes anisotropic, so the net spin current carried by the scattered electrons does \emph{not} vanish after summation over all $\bvec{k}$-states (Fig.~\ref{From2Dto3D}(d)). The spin current carried by the scattered electrons flows out-of-plane with net spin direction along $\zhat \times \bvec{E}$. Following Ref.~\onlinecite{iSOCAminiGSC}, we refer to this mechanism of generating spin currents as \emph{spin-orbit filtering}, because the Rashba spin-orbit potential filters the unpolarized, incident spins as they scatter off the interface, yielding an out-of-plane spin current. Note that while the spin Hall effect occurs in bulk materials and spin-orbit filtering occurs only at interfaces, both effects can generate spin currents with the same spin and flow orientation, making them difficult to distinguish in experiments. Unlike the spin Hall effect, which mostly depends on material properties from the single originating layer, spin-orbit filtering depends strongly on the momentum relaxation times and electronic structure of the two adjacent layers and requires inversion symmetry to be broken by the interface.

If one of the layers is ferromagnetic, there is another interfacial mechanism that generates spin currents. Assume that in one layer the in-plane charge current is spin-polarized along $\bvec{p}$. This occurs in ferromagnetic layers, where $\bvec{p}$ points along the magnetization $\mhat$. In this case, the spin polarized carriers will rotate about the spin-orbit field while scattering off the interface, as seen in Fig.~\ref{From2Dto3D}(e). This phenomenon occurs in addition to the filtering effect described above. After summing over all $\bvec{k}$-states, the net out-of-plane spin current  has a component with the spin direction along $\bvec{p} \times (\zhat \times \bvec{E})$ (Fig.~\ref{From2Dto3D}(f)). We refer to this phenomena as \emph{spin-orbit precession}, because it describes spins precessing about the spin-orbit field while they scatter off the interface. The spin swapping effect, first predicted in Ref.~\cite{SSTheoryLifshitsDyakonov}, has a similar phenomenogical form to spin-orbit precession when it occurs near interfaces\cite{SSTheorySaidaoui,SSTheoryPauyac}. Like the spin Hall effect and spin-orbit filtering effects, spin swapping and spin-orbit precession differ in that the latter depends more intimately on the relaxation times and electronic structure of both material layers. Another key difference is that the flow and spin orientations described by spin swapping represent a subset of those allowed by the spin-orbit precession mechanism, as discussed in section \ref{sec:mechanisms}.

The spin-orbit filtering and precession effects discussed here represent the simplest generalization of the Rashba-Edelstein model\cite{SOTTheoryManchon} that sparked this field of study. However, several other important spin-orbit torque mechanisms have been predicted. For instance, theory predicts spin-orbit torques that are directly generated at interfaces\cite{kurebayashi2014antidamping} that share a common origin with the spin Hall effect but are not caused by that effect. The spin Hall effect arises due to both intrinsic and extrinsic mechanisms, which we discuss below. The intrinsic mechanism can be interpreted as capturing the perturbation of the electronic wavefunctions under an applied electric field, creating nonequilibrium electronic states that carry spin currents. The same perturbation to electronic wavefunctions occurs for carriers in regions that break inversion symmetry (like at interfaces), yielding the additional torques that were proposed in Ref.~\onlinecite{kurebayashi2014antidamping}.

\subsection{In-plane Transport in Trilayers}
\label{sec:InPlaneTrilayers}

Trilayers have a more complex geometry than bilayers, and therefore have more degrees of freedom and experimentally-controllable (and uncontrollable) parameters. For instance, a spin valve is a trilayer system where the magnetization directions of each ferromagnetic layer can be different. Based on symmetry considerations alone, the magnetization dependencies of the torques in trilayers are more complex than in bilayers. This additional complexity occurs in part because spin currents originating in one ferromagnetic layer or at the adjacent interface can flow through the spacer layer and exert torques on the other ferromagnetic layer. By varying each ferromagnetic layer's magnetization direction and/or selectively inserting additional layers, one can obtain information about spin-orbit torques not available in bilayers that could help parse spin-orbit torque mechanisms. As a result, trilayers present unique structures to further investigate the spin-orbit torques first proposed in bilayers.

\begin{figure}
	\centering
	\vspace{0pt}	
	\includegraphics[width=1\linewidth,trim={0.0cm 0.0cm 0.0cm 0.0cm},clip]{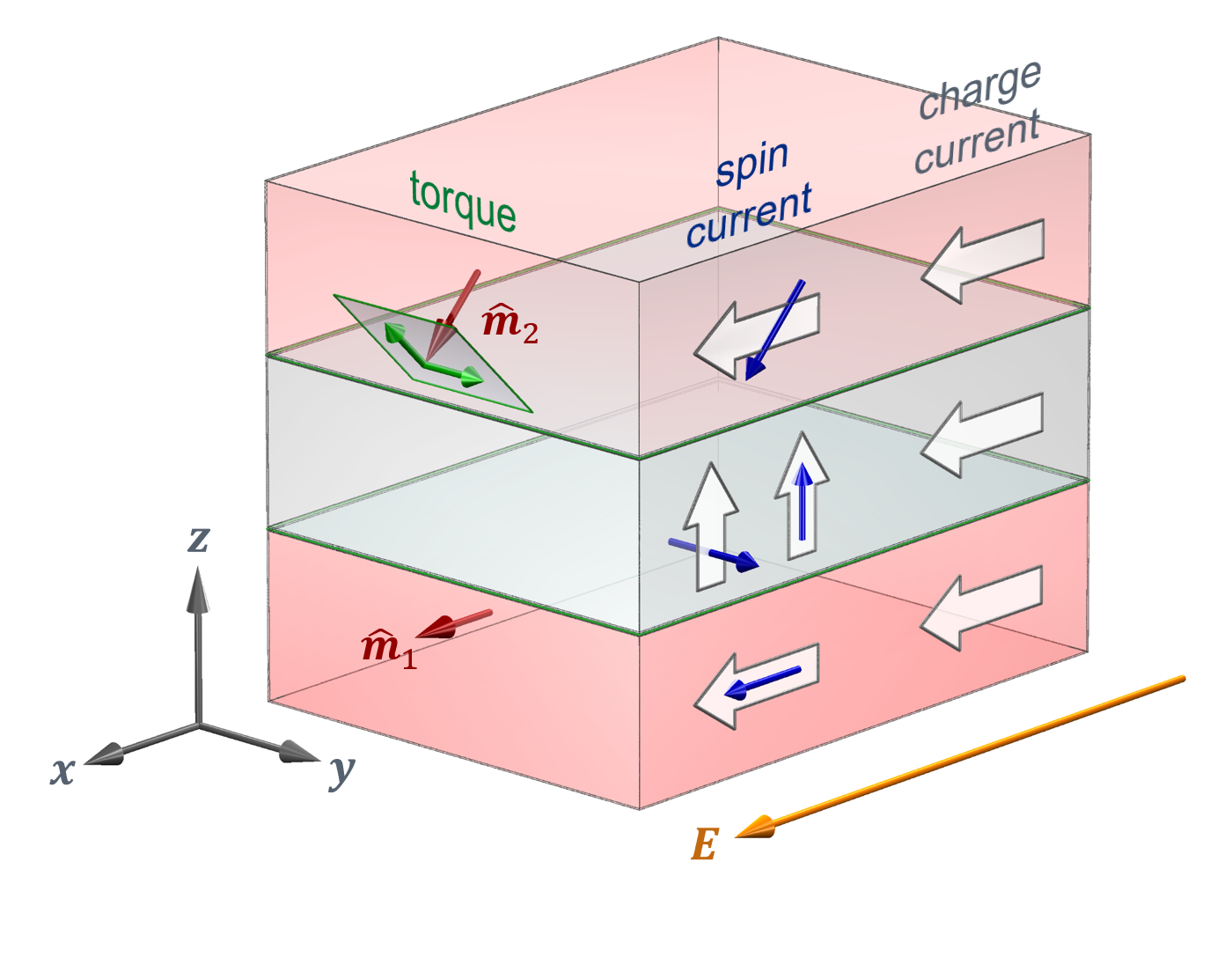}
	\caption{Magnetic trilayer and in-plane transport. The top and bottom ferromagnetic layers are separated by a non-magnetic layer.  In each layer, the current flows along the electric field, where flow directions are given as block arrows. In the ferromagnetic layers, the spin currents shown here flow along the charge currents with spins aligned with the magnetization (indicated by red arrows). Note that for spin currents, block arrows give flow direction and blue arrows give spin direction. Green arrows indicate the two components of the torque on the magnetization. In the non-magnetic layer, spin currents originating in the lower ferromagnetic layers and flowing out-of-plane ($\zhat$) have spin directions along both $\yhat$ and $\zhat$ (other contributions to the spin current are not shown). The spin currents with $y$-spin direction can arise from the spin Hall effect in any layer or through the spin-orbit filtering effect at interfaces. The spin currents with $z$-spin direction are not allowed by symmetry in bulk nonmagnets; these spin currents can only arise in the ferromagnetic layers through various processes or at the interfaces through the spin-orbit precession effect. Spin transfer torques arising from the spin currents with $z$-spin direction could switch ferromagnetic layers with perpendicular magnetic anisotropy, suggesting applications of possible technological interest.
	}
	\vspace{0pt}
	\label{TrilayerSOT}
\end{figure}

Measurements of spin-orbit torque in trilayers are nearly as old as measurements in bilayers. Trilayers were first investigated as means to suppress interfacial spin-orbit torque contributions.\cite{fan2013observation,fan2014quantifying,nan2015comparison} In these experiments, a light spacer material (typically Cu) with a spin diffusion length far exceeding layer thicknesses was placed in between the heavy metal and the ferromagnet, resulting in a trilayer. Since the heavy metal is no longer adjacent to the ferromagnet, no interface exists in the trilayer that has both strong spin-orbit coupling and ferromagnetic exchange. The lack of such an interface was thought to suppress interfacial spin-orbit torques, such that all torques could be attributed to spin current generation in the heavy metal.

Two problems exist with this interpretation. First, those trilayers still contain a nomagnetic interface with strong spin-orbit coupling (formed by the heavy metal and the spacer layer). As previously discussed, under an in-plane electric field, theory predicts that these interfaces can generate spin currents of comparable strength to spin Hall currents in Pt. If such interface-generated spin currents occur in the system, the measured torque is no longer solely due to the spin Hall effect in the heavy metal. Second, despite lower spin-orbit coupling strength at the interface between the ferromagnet and spacer layer, both interfacial and anomalous spin-orbit torques could still contribute to the measured torques.

The trilayers discussed so far have a single ferromagnetic layer. Investigations of trilayers with two ferromagnetic layers and a nonmagnetic spacer layer (spin valves) have expanded the reach of spin-orbit torque measurements \cite{SOTExpHumphries, iSOCBaekAmin}. In these experiments, the spin currents generated in ferromagnets or at adjacent interfaces can be measured through their effect on the other ferromagnetic layer, creating an independent measurement of spin-current driven torques. In the following, we discuss the ramifications of spin currents generated both at interfaces and in bulk ferromagnetic layers on spin torques in trilayers.

\emph{Interfacial spin current generation in trilayers}---The experimental results reported in Ref.~\onlinecite{iSOCBaekAmin} demonstrated that if one ferromagnetic layer has an out-of-plane magnetization, current-induced torques could switch that magnetization without external magnetic fields if the other ferromagnetic layer's magnetization was in-plane. This behavior is allowed by symmetry, and one possible mechanism explaining these results involves interface-generated spin currents. The spin-orbit precession current generated at the interface between the in-plane magnetized layer and the spacer layer has spin direction $\mhat_\text{IP} \times (\zhat \times \bvec{E})$, where $\mhat_\text{IP}$ is the in-plane magnetization direction. If the electric field and in-plane magnetization are parallel, the resulting out-of-plane spin current has an out-of-plane spin direction. This spin current can then flow through the spacer layer into the out-of-plane ferromagnetic layer and exert a spin transfer torque with the right orientation to enable switching. The authors presented evidence of spin currents with out-of-plane spin direction in the form of this field-free switching, and through current-induced shifting of the hysteresis loops of the out-of-plane layer.

Recent experiments have expanded these findings by considering different magnetization configurations and using other experimental techniques. Hibino et al.\cite{iSOCHibino} investigated spin-orbit torques in Py/Pt/Co trilayers using harmonic Hall analysis. They found two distinct dampinglike torques through the angular dependence of the harmonic Hall signal, which damped the magnetization towards the $\bvec{p}$ and $\zhat \times \bvec{p}$ directions, where $\bvec{p} = \zhat \times \bvec{E}$. Spin transfer torques originating from the spin Hall effect damp the magnetization towards $\bvec{p}$ and can incite magnetization precession about $\bvec{p}$ through spin-dependent scattering at the interface (parameterized by the spin mixing conductance). However, torques that damp the magnetization towards the $\zhat \times \bvec{p}$ cannot be explained by the spin Hall effect since its spin direction (which points along $\bvec{p}$) is tightly constrained by symmetry. The authors attributed the unconventional dampinglike torque to the spin-orbit precession effect at the Pt/Co interface and extracted an associated spin torque efficiency in reasonable agreement with first principles calculations.\cite{iSOCAminiGSC} The authors further showed that the spin torque strength depended greatly on the material composition of the interfaces, further suggesting an interfacial origin. In another work,\cite{iSOCHibino2} Hibino et al.~find further experimental evidence of spin currents with both $\bvec{p}$ and $\zhat \times \bvec{p}$ spin directions generated in FeB/Cu/CoNi multilayers, this time using spin torque ferromagnetic resonance techniques to measure the angular dependence of the associated torques.

\emph{Bulk ferromagnetic spin current generation in trilayers}---Many experiments have measured spin currents originating in ferromagnetic layers with a magnetization-aligned spin direction.\cite{FSHEExpWu,FSHEExpGibbons,FSHEExpBose,FSHEOmori} Since charge currents in ferromagnets are spin-polarized, both the planar and anomalous Hall effects are expected to generate spin-polarized currents with spin directions aligned with the magnetization \cite{AHEAMRTaniguchi,safranski2020planar}, which could explain some of these recent experiments. Other experiments  measured contributions from both transverse and magnetization-aligned spin directions in Py,\cite{FSHEExpMiao,FSHEExpTian,FSHEExpDas} further supporting the claim that ferromagnets are robust generators of spin current. While magnetization-aligned spin currents cannot exert spin torques in single layer ferromagnets, they can exert torques on other ferromagnetic layers within trilayers, as long as the magnetization direction of the other ferromagnetic layer is noncollinear to the magnetization of the generating layer.

\subsection{Extrinsic vs Intrinsic Effects}
\label{sec:extint}

The electrical control of magnetization through spin-orbit torques can be understood by examining the response of the system, Eq.~\ref{eq:Hgeneral}, to an applied electric field.  Heavy metal-ferromagnet thin film bilayers typically operate in the linear response regime.  In this case, the electric field impacts the system in two ways: first by changing the electrons' distribution function, and second by changing the electrons' wave functions.  Often, each of these two aspects of the electric field perturbation result in the same observable (e.g. anomalous Hall current).  The prefix ``extrinsic'' or ``intrinsic'' indicates the physical mechanism under consideration ({\it e.g.}, extrinsic versus intrinsic anomalous Hall current).  While there is not a universally agreed upon usage of intrinsic and extrinsic, we find it useful to use ``extrinsic'' to describe the contributions where the perturbing electric field changes the occupation of the states and ``intrinsic'' to describe the contributions where the perturbing electric field changes the states themselves.  This distinction is straightforward to make in calculations based on the Kubo formula but may not be so straightforward in other approaches.

When the electric field perturbs the distribution function, the nonequilibrium distribution function can be obtained by solving the Boltzmann equation, and has so far been studied in the relaxation time approximation.\cite{SOTTheoryManchon,SOTTheoryHaney,garate2009influence,SOTTheoryFreimuth} Scattering plays a central role in determining the nonequilibrium distribution function and all subsequent observables (e.g. charge and spin current, magnetization torques).  The magnitude of these effects typically scale linearly with the scattering time $\tau$, in the limit where $\hbar/\tau < \varepsilon$, where $\varepsilon$ is an energy scale that is characteristic of the typical band splitting near the Fermi energy. Interestingly, this scaling implies that when scattering is very weak ({\it e.g.}, $\tau$ is very large) the scattering-based contribution to the spin Hall conductivity, for example, dominates over other contributions.\cite{zhu2019variation} A more common regime for transition metals is the clean to dirty metal limit, in which the intrinsic mechanism dominates. In this case, the intrinsic response is independent of $\tau$ for $\hbar/\tau < \varepsilon$,\cite{tanaka2008intrinsic} and varies as $\tau^m$ for $\hbar/\tau < \varepsilon$, where $m=2$ for simple models of scattering,\cite{tanaka2008intrinsic} but whose specific value generally depends on the observable and the microscopic model.\cite{AHEReviewNagosa}

The extrinsic and intrinsic contributions have been studied extensively for the two-dimensional Rashba model, Eq.~\ref{eq:Hrashba}.  The extrinsic response was analyzed in Refs.~\onlinecite{manchon2008theory, manchon2009theory}.  The application of an electric field perturbs the distribution function, introducing asymmetry in the occupation of states with Bloch wave vector along $\bm{E}$. This naturally leads to a spin accumulation aligned in the $\bm{E}\times\bm{z}$ direction, as illustrated in Fig.~\ref{From2Dto3D}(b).  In the remainder of this review, we focus on extrinsic contributions to the interfacial spin-orbit torque.

\subsection{Symmetry}
\label{sec:symmetry}
  
In this section, we demonstrate how symmetry constrains spin-orbit torques in various material systems following earlier discussions.\cite{garello2013symmetry, belashchenko2019first} We first show that in nonmagnet/ferromagnet bilayers, symmetry forces the torque to vanish along the axis $\bm{\hat{z}} \times \bm{E}$. Next, we derive the general form of the response tensor that relates the torque and the electric field as a function of magnetization direction, first considering only continuous
symmetries and later including discrete crystal symmetries. Finally, we discuss how unique crystal symmetries affect spin-orbit torque and lead to novel phenomena.

\begin{figure}[b!]
	\centering
	\vspace{0pt}	
	\includegraphics[width=0.95\linewidth,trim={0.0cm 1.2cm 0.0cm 0.0cm},clip]{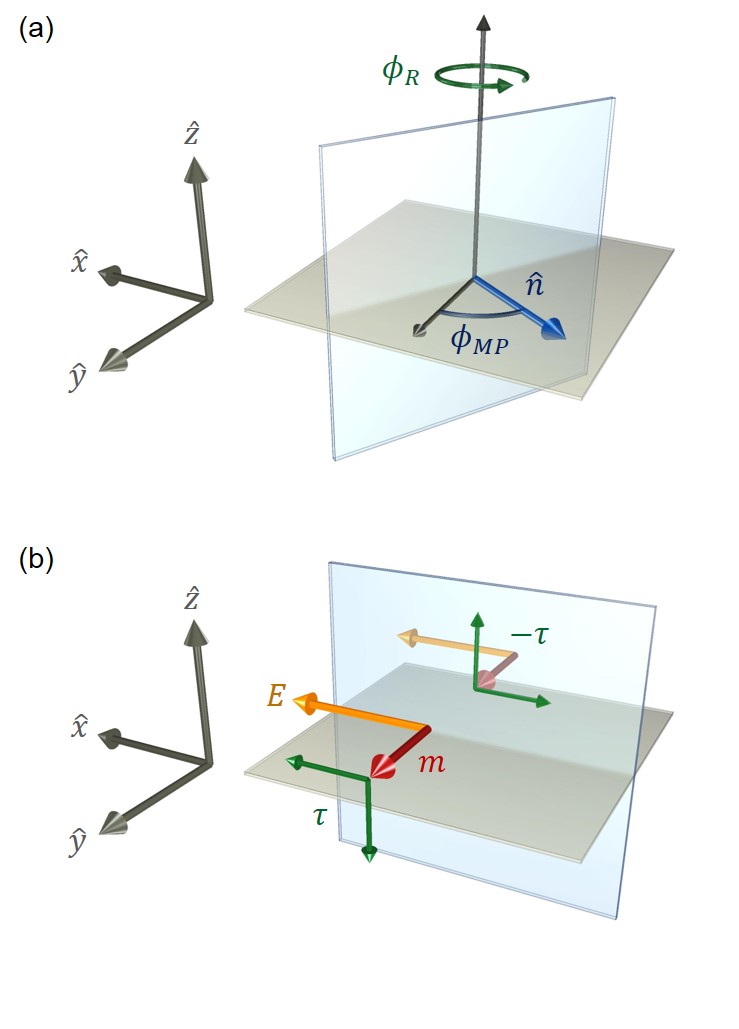}
	\caption{
	Depiction of symmetries and their consequences in a polycrystalline bilayer. (a) For nonmagnetic bilayers, any rotation $\phi_\text{R}$ about the out-of-plane direction ($z$-axis) leaves the system unchanged. Likewise, any mirror-plane transformation where the mirror-plane normal lies in-plane (parameterized by the angle $\phi_\text{MP}$) also leaves the system invariant. (b) Under an applied, in-plane electric field $\bvec{E}$, all symmetries are broken except the mirror-plane that lies parallel to the electric field, since the electric field is a polar vector.
	If one layer is ferromagnetic, this symmetry is broken unless the magnetization $\mhat$ points normal to the mirror-plane, since magnetization is a pseudovector. In this configuration, the torque $\bvec{\tau}$ must vanish, because a nonvanishing torque reflected through the mirror-plane will reverse, violating the system's symmetry.
	}
	\vspace{0pt}
	\label{MirrorPlaneSymmetry}
\end{figure}

Ignoring crystal structure, only two types of spatial symmetries exist
in a nonmagnetic bilayer: 1) continuous rotational symmetry about
$\zhat$ and 2) mirror-plane symmetry with respect to planes whose normal vector $\nhat$ lies within the interface plane. \Fig{MirrorPlaneSymmetry}a illustrates
these symmetries, where $\phiR$ denotes the angle of rotation about
$\zhat$ and $\phiMP$ denotes the angle of the mirror-plane normal
vector $\nhat$. As we show, $\phiR$ and $\phiMP$ parameterize
every spatial symmetry transformation for the bilayer system, where
$\phiR \in [0,2\pi]$ and $\phiMP \in [0,\pi]$.

An applied, in-plane electric field $\bvec{E}$ breaks all of these
symmetries except the single mirror-plane that contains $\bvec{E}$
(i.e. when $\nhat \perp \bvec{E}$). To see this, note that $\bvec{E}$
is a polar vector, so when $\bvec{E}$ lies within the mirror-plane it
is invariant upon reflection (\Fig{MirrorPlaneSymmetry}b). All
rotations about $\zhat$ will change the orientation of $\bvec{E}$
since it lies in-plane, so those transformations are no longer
symmetries of the system. Assuming now that one layer is
ferromagnetic, its magnetization direction $\mhat$ must also be
invariant to any allowed symmetry transformation. Since magnetization is a
pseudovector, it is invariant to a mirror-plane reflection only when
it is normal to the mirror-plane. Thus, as depicted in
\Fig{MirrorPlaneSymmetry}b, the only remaining symmetry is a mirror
plane containing $\bvec{E}$ and $\zhat$ when $\mhat = \pm \zhat \times
\bvec{E}$. Only in this scenario is $\bvec{E}$ within the mirror-plane
and $\mhat$ normal to the mirror-plane.

Under the assumption that only the magnetization direction is variable
(i.e. the magnetization is saturated), the net torque acting on the magnetization must be
orthogonal to $\mhat$. Thus, for the case when $\mhat$ points normal
to the mirror-plane as described above, the torques $(\tau)$ must be
parallel to the mirror-plane. Since torque is a pseudovector, and
since any pseudovector parallel to a mirror-plane will flip sign upon
reflection (see \Fig{MirrorPlaneSymmetry}b), the torque must vanish to
preserve the mirror-plane symmetry. Thus, in nonmagnet/ferromagnet
bilayers under an applied, in-plane electric field, all spin-orbit
torques must vanish when $\mhat = \zhat \times \bvec{E}$.

\begin{figure*}[t!]
	\centering
	\vspace{0pt}
	\includegraphics[width=0.9\linewidth,trim={0.0cm 1.0cm 0.0cm 0.0cm},clip]{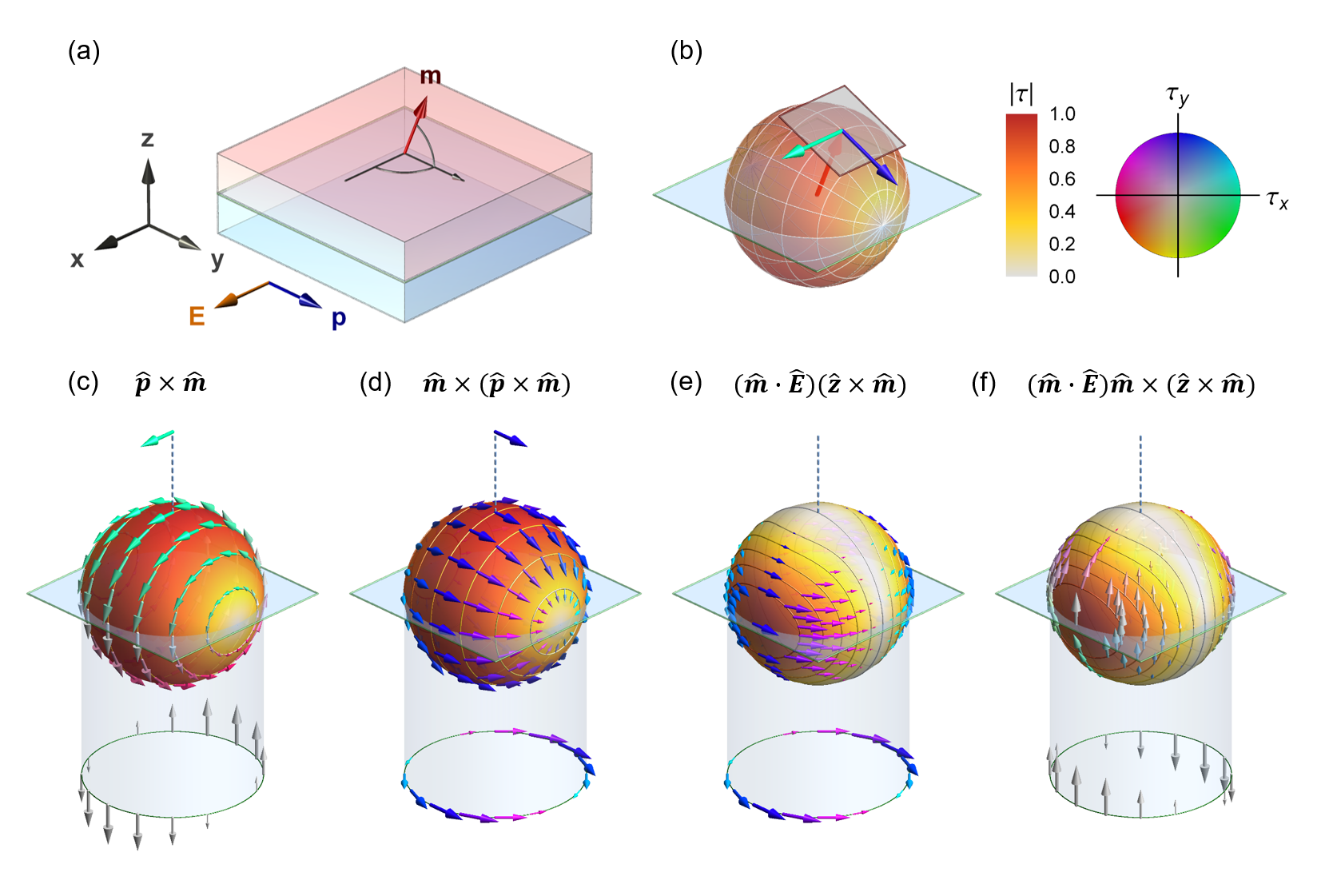}
	\caption{
	(a) Depiction of a bilayer consisting of a heavy metal layer (blue region) and a ferromagnetic layer (red region) under an applied, in-plane electric field $\bvec{E}$ (here along $\xhat$). The high symmetry direction $\bvec{p} = \zhat \times \bvec{E}$ is normal to the $x/z$ mirror-plane, where $\zhat$ points out-of-plane. (b) Spin-orbit torque is conveniently defined using two basis vectors: dampinglike ($\mhat \times ( \bvec{p} \times \mhat )$) and fieldlike ($\bvec{p} \times \mhat$), which are defined relative to the high-symmetry direction $\bvec{p}$. These basis vectors span the plane perpendicular to the magnetization and vanish when $\mhat~||~\bvec{p}$, satisfying the bilayer's symmetry constraints. However, the dampinglike and fieldlike basis vectors are not sufficient to describe the magnetization-dependence of spin-orbit torque unless they have magnetization-dependent coefficients. The full expansion of spin-orbit torque using constant coefficients is more complicated, and is given by \Eq{FullExp}. This expansion consists of four-vector functions of the magnetization, depicted in (c)-(f). Each vector function can be additionally multiplied by any power of $m_z^2$. The in-plane and out-of-plane torques are projected below and above the unit sphere respectively. The full expansion suggests that if measurements of in-plane and out-of-plane torques are interpreted as arising from only dampinglike or fieldlike torques, the full magnetization dependence may be misrepresented. For example, when $\mhat~||~\zhat$, measuring a small torque component pointing along $\bvec{p}$ indicates a small dampinglike torque, but this measurement could be incorrectly interpreted as weak potential for magnetization switching, because the torque component $(\mhat \cdot \bvec{E}) \zhat \times \mhat$ shown in (e) is zero at $\mhat~||~\zhat$ but contributes to the switching process for other magnetization directions.
	}
	\vspace{0pt}
	\label{SymTorques}
\end{figure*}

So far, we have considered two constraints on spin-orbit torques in
nonmagnet/ferromagnet bilayers: 1) they point orthogonally to $\mhat$
and 2) they vanish when $\mhat = \zhat \times \bvec{E}$. Given these
constraints, spin-orbit torques could be written as a linear
combination of the following terms,
\begin{align}
\bvec{\tau}_D	&=  c_D(\mhat)		\mhat \times \big{(}	\bvec{p} \times \mhat	\big{)}		\label{DLT}	\\
\bvec{\tau}_F		&=  c_F(\mhat)								\bvec{p} \times \mhat	,		\label{FLT}
\end{align}
where $D$/$F$ refers to the dampinglike/fieldlike component
(visualized in \Fig{SymTorques}), $c_{D/F}(\mhat)$ are
magnetization-dependent scalar functions, and $\bvec{p} = \zhat \times
\Ehat$. The dampinglike and fieldlike vectors span the plane
perpendicular to the magnetization and vanish when $\mhat = \bvec{p}$,
satisfying the two constraints above. \Eqs{DLT}{FLT} describe two
types of behavior: damping towards $\zhat \times \Ehat$ (\Eq{DLT})
and precession about $\zhat \times \Ehat$ (\Eq{FLT}).

It is important to note that, in general, the coefficients
$c_{D/F}(\mhat)$ cannot be given by all functions of $\mhat$.  In
other words, \Eqs{DLT}{FLT} are under-constrained. The general expression\cite{garello2013symmetry} for spin-orbit torque in a bilayer subject to the continuous symmetries described above is given by
\begin{align}
\bvec{\tau} =  ~& \sum^\infty_{l=0} (m_z)^{2l} \big{(} 
                a_{l} \bvec{p} \times \mhat + 
                b_{l} \mhat \times (\bvec{p} \times \mhat) \nonumber \\
                &+c_{l} (\mhat \cdot \bvec{E}) \zhat \times \mhat +
                d_{l} (\mhat \cdot \bvec{E}) \mhat \times (\zhat \times \mhat) \big{)}
                \label{FullExp}
\end{align}
where $a_l$, $b_l$, $c_l$, and $d_l$ are the coefficients of expansion. Note that the first four terms in the expansion (zeroth order in $m_z$) are the traditional fieldlike torque and dampinglike torque plus two additional terms. The additional terms behave like fieldlike and dampinglike torques defined relative to $\zhat$ instead of $\bvec{p} = \zhat \times \Ehat$, but also carry a factor $\mhat \cdot \Ehat$, which ensures the torque vanishes when $\mhat = \bvec{p}$ as required.

The vector forms in Eq.~\ref{FullExp} are shown in Fig.~\ref{SymTorques}(c-f).  Each of these forms can additionally be multiplied by $(m_z)^{2l}$, each power of which suppresses the torque at $\theta=0,\pi$. An important point is that measuring the torque at the poles, $\theta=0,\pi$, or the equator, $\theta=\pi/2$ does not necessarily predict the behavior at the other set of points. The difference, if large, can be important for connecting measurements of the torque at specific magnetization directions with magnetic dynamics, which depend on the values of the torques at many points.\cite{garello2013symmetry} There are indications in both model calculations\cite{lee2015angular} and first principles calculations\cite{belashchenko2019first} that the angular dependence can be more complicated than just a sum of the simple fieldlike and dampinglike torques. The expansion in Eq.~\ref{FullExp} is complete and it is easy to envision what each of the terms looks like.  Unfortunately, the different terms are not orthogonal to each other.  That means that if a finite truncation of the series is used to fit experimental (or calculated) data, the fit parameters, $a_{l}$ etc., will depend on the order at which the series is truncated.  Ref.~\onlinecite{belashchenko2020interfacial} gives an orthogonal expansion in terms of modified vector spherical harmonics. That expansion is more appropriate for fitting data although the higher order terms have a less transparent form.

We finally note that, quite generally, reducing the system symmetry relaxes the constraints on the system response and enables more nonzero components of the torque.  Recent work has utilized substrates which lack in-plane mirror symmetry, such as transition metal dichalcogenides WTe$_2$, MoTe$_2$, and others.\cite{macneill2017control,macneill2017thickness,guimaraes2018spin,stiehl2019layer,shi2019all,shao2016strong}  For these materials, applying a current in the single mirror-plane results in an out-of-plane torque. Such a torque may enable switching of perpendicular magnetic layers, which possess technological advantages relative to in-plane magnetic layers.\cite{wang2013low,garello2014ultrafast}

\subsection{Distinguishing Bulk from Interface Effects}
\label{sec:lengthScales}

For spin transfer torques and current-perpendicular-to-the-plane giant magnetoresistance,\cite{bass1999current} the most important length scale is the spin diffusion length, the length scale over which spin currents and spin accumulations decay.  It is defined as the distance a spin diffuses before it undergoes spin-flip scattering and is given by $\ell_\text{sf}^2=\lambda v_{F} \tau_\text{sf} / 6$, where $\lambda$ is the elastic mean free path, which is the average distance between elastic scattering events, $v_{F}$ is the Fermi velocity, and $\tau_\text{sf}$ is the spin-flip scattering time.  On the other hand, for current-in-the-plane giant magnetoresistance, the important length scale is the mean free path.  Spin-orbit torques combine both in-plane charge transport with out-of-plane spin transport making it likely that both length scales are important.  The importance of multiple length scales can complicate the interpretation of calculations and experiments.

Other length scales may be important as well.  For example, the spin current associated with the spin Hall effect differs qualitatively from diffusive spin currents.  If the spin Hall spin current is intrinsic (see Sec.~\ref{sec:extint}), it can be described as arising from an anomalous velocity of electrons at special points on the Fermi surface where spin-orbit coupling leads to large band splittings.  It is not clear whether such spin currents vary with the spin diffusion length  or with yet a different length scale.  Studies of the thickness dependence\cite{kim2013layer,SPExpSanchez,nguyen2016spin} are frequently interpreted by assigning the observed length scale to the spin diffusion length, but that length scale could be entirely different.

A final unknown length scale that complicates the interpretation of experiment is the length scale over which structural details of the layers vary.  There are many processes that can contribute to structural variations, including relaxation of strain due to lattice mismatch and grain growth.  Without detailed structural characterization and related calculations it is difficult to know how much of a measured variation with layer thickness could be due to structural changes.
 
  \section{Phenomenological Mechanisms}
  \label{sec:mechanisms}

Building on the idea that the overlapping spin-orbit coupling and exchange interaction at an interface can make a substantial contribution to the spin-orbit torque, we introduce an extremely simple model that displays all of the important interfacial effects. 

First, we describe a spin torque on a strong magnetic impurity, where the absorbed spin current equals the spin torque in the absence of spin-orbit coupling. We then show that in the presence of spin-orbit coupling this is no longer the case, because spin-orbit coupling opens another channel of angular momentum transfer to and from the atomic lattice. 

Next, we introduce a model based on embedding a two-dimensional Hamiltonian describing the interface within a structureless, three-dimensional bulk material, yielding a bilayer. The two-dimensional Hamiltonian describing the interface breaks structural inversion symmetry by design, enabling a nonvanishing spin-orbit torque. In this approach, the important interactions are due to electrons reflecting from or transmitting through the interface.  As they do, they interact with the interfacial spin-orbit coupling and exchange interaction, which modify the transmission and reflection amplitudes.  These amplitudes combine to give all of the currents and spin currents at the interface as well as the torques on the magnetization.

The simple models presented here are not realistic since details of the electronic structure and disorder are absent. However, such models accomplish three goals: 1) they give a qualitative understanding of interfacial spin-orbit effects, 2) they present a quasi-analytical way of separating contributions from spin-orbit coupling and the exchange interaction, and 3) they provide a template to compare to ab-initio calculations so that some physical intuition may be extracted.

\subsection{Spin torque on a strong magnetic impurity in one dimension}
\label{subsec:SpinTorque}

In this section we provide a simple example of a spin torque on a strong magnetic impurity. Later, we show that the qualitative behavior in this example is analogous to the role that interfaces play on spin torques in bilayers.

\begin{figure}
	\centering
	\vspace{0pt}	
	\includegraphics[width=1\linewidth,trim={0.0cm 0.0cm 0.0cm 0.0cm},clip]{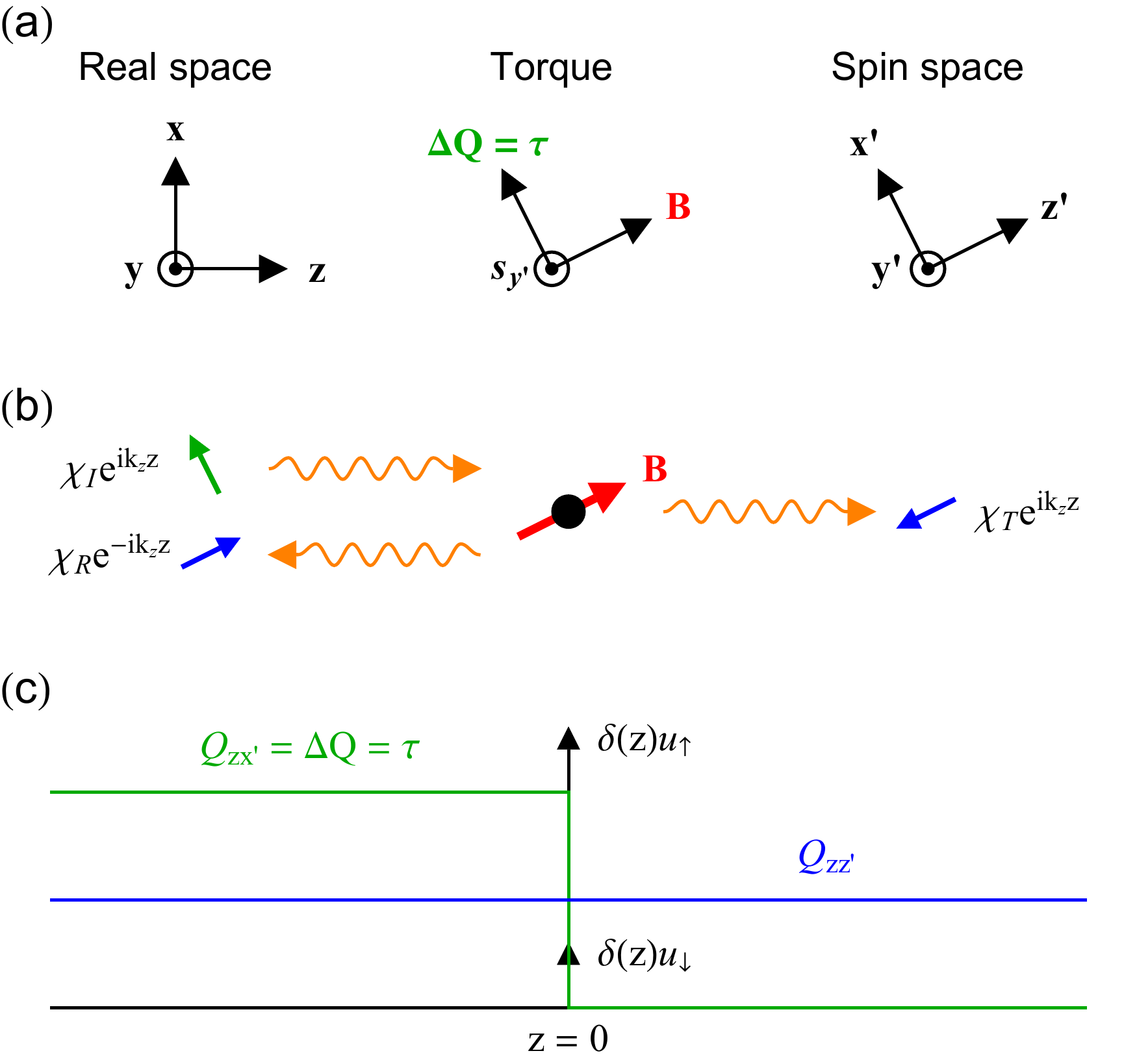}
	\caption{
	Schematic of an electron scattering off of an infinitely strong magnetic impurity. (a) Coordinate systems for real space and spin space, where transport occurs along $z$ and the impurity's magnetic moment points along $z'$. The basis states $\ket{\UA}$ and $\ket{\DA}$ correspond to spins along $\pm z'$. (b) In the limit that $\bvec{B} \rightarrow \infty$, the impurity perfectly reflects $\UA$ spins and perfectly transmits $\DA$ spins. Thus, for an incoming spin state $\chi_\inc \propto \ket{\UA} + \ket{\DA}$ along $x'$ (transverse to the impurity's magnetic moment), the reflected and transmitted spin states point along $\pm z'$ (parallel or antiparallel to the impurity's magnetic moment) (c) Plot of the spin current as a function of position. The incoming state carries spin current $Q_{zx'}$; the indices indicate flow along $z$ and spin direction along $x'$. The reflected and transmitted states each carry the same spin current $Q_{zz'}$; the indices indicate flow along $z$ and spin direction along $z'$. Thus, the net change in spin current across the impurity is $Q_{zx'}$, indicating that the incoming spin angular momentum is completely absorbed. The absorption of spin current results in a torque on the impurity's magnetic moment.
	}
	\vspace{0pt}
	\label{ScatProb1D}
\end{figure}

Imagine an electron scattering off of a magnetic impurity in one dimension, described by the coordinate $z$, (Fig. \ref{ScatProb1D}). We assume the magnetic impurity (located at $z = 0$) is captured by a delta function potential
\begin{align}
    V_{\UD}(z)   &\propto \delta(z) \uUD,
    \label{pot}
\end{align}
where $\uUD$ is the barrier strength for spins parallel ($\UA$) or antiparallel ($\DA$) to the magnetic field of the impurity ($\bvec{B}$). In what follows, we use the coordinate system $(x',y',z')$ for spin space, such that the $z'$ direction points along $\bvec{B}$.

For simplicity, let $\uD = 0$ and $\uU \rightarrow \infty$. The impurity thus behaves as a perfect reflector for $\UA$ spins and a perfect transmitter for $\DA$ spins, though each spin state can acquire a phase factor upon scattering. Now assume an electron with spin transverse to $\bvec{B}$ scatters off of this impurity. The incoming transverse spin state could be described by the following spinor (assuming spin along $x'$):
\begin{align}
    \chi_\inc = \frac{1}{\sqrt{2}} \big{(} \ket{\UA} + \ket{\DA} \big{)}.
\end{align}
The reflected and transmitted states are then
\begin{align}
    \chi_\refl = \frac{\alpha}{\sqrt{2}}\ket{\UA}, \quad \chi_\trans = \frac{\beta}{\sqrt{2}}\ket{\DA},
\end{align}
where $\alpha$ and $\beta$ are the phase factors acquired upon scattering. Thus, while the incoming spin is transverse to $\bvec{B}$ (along $x'$), the reflected and transmitted spins are parallel and antiparallel to $\bvec{B}$ (along $\pm z'$). To ensure continuity of the wavefunction at $z = 0$, such that $\chi_\inc + \chi_\refl = \chi_\trans$, we may choose $\alpha = -1$ and $\beta = 1$. 

The incident, reflected, and transmitted states each carry a spin current that flows along $z$ (Fig. \ref{ScatProb1D}b/c). Since the reflected and transmitted states have equal and opposite spin \emph{and} equal and opposite velocities, they both carry the same spin current ($Q_{zz'}$), which is continuous across the impurity and implies no angular momentum transfer. However, the incoming spin current ($Q_{zx'}$) is completely absorbed by the impurity, since this spin current exists on one side of the interface but not on the other side. The absorption of the incoming spin current is a simple example of a spin transfer torque ($\bvec{\tau}$), where in this case $\bvec{\tau} = Q_{zx'}\xhat'$.

For there to be a spin transfer torque along $x'$ on the impurity, the spin density $\bvec{s}$ at $z = 0$ must have a non-vanishing $y'$ component and a vanishing $x'$ component. This is because the spin torque is given in general by:
\begin{align}
    \bvec{\tau} \propto (\uU - \uD) \bvec{s} \times \bvec{B} \rightarrow \uU \bvec{s} \times \bvec{B},
    \label{torque}
\end{align}
where the last form holds for $\uD=0$.  Since $\bvec{\tau}~||~\xhat'$ and $\bvec{B}~||~\zhat'$, the spin accumulation $\bvec{s}$ must lie in the $y/z$ plane to satisfy \Eq{torque}. Here we run into an apparent problem. If the wavefunction at $z = 0$ equals $\chi_0 = \chi_\inc + \chi_\refl = \chi_\trans = \ket{\DA}$, then the spin density $\bvec{s}$ points solely along $-z'$ and has no $y'$ component, resulting in a vanishing torque. Where then did the angular momentum from the absorbed, incoming spin current go?

The inconsistency described above is corrected by carefully taking the limit as $\uU \rightarrow 
\infty$. As we show in Appendix~\ref{appendix:spintorque}, the wavefunction at $z = 0$ is actually a superposition of spin states given by
\begin{align}
    \chi_0 = \chi_\trans \propto -(\im a/\uU)\ket{\UA} + \ket{\DA},
\end{align}
because when $\uU$ is large but not infinite, a tiny amount of $\ket{\UA}$ must be transmitted (here we have omitted the normalization factor). Note that $a$ is a dimensionless constant determined by details of the scattering potential. Then, as $\uU \rightarrow \infty$, $\chi_0$ approaches the previous solution, but now carries a component of spin along $y'$ as required
\begin{align}
    \bvec{s} &= \chi_0^\dagger \bvec{\sigma} \chi_0 = 
            \begin{pmatrix} 0 \\ 2a/\uU \\ -1 + a/\uU^2
            \end{pmatrix}
            \rightarrow
            \begin{pmatrix} 0 \\ 0 \\ -1 \end{pmatrix}
    \label{s0vec}
\end{align}
where $\bvec{\sigma}$ is the vector of Pauli matrices. Even though $s_{y'}$ vanishes as $\uU \rightarrow \infty$, the torque it exerts does not, because the prefactor $\uU$ in \Eq{torque} exactly cancels the factor of $1/\uU$ in $s_{y'}$. Because the torque equals the incoming spin current, which does not depend on $\uU$, the torque cannot depend on $\uU$ either. The cancellation of $\uU$ reflects this fact.


\subsection{Role of spin-orbit coupling on the torque}

\begin{figure}[t!]
	\centering
	\vspace{0pt}	
	\includegraphics[width=1\linewidth,trim={0.0cm 0.0cm 0.0cm 0.0cm},clip]{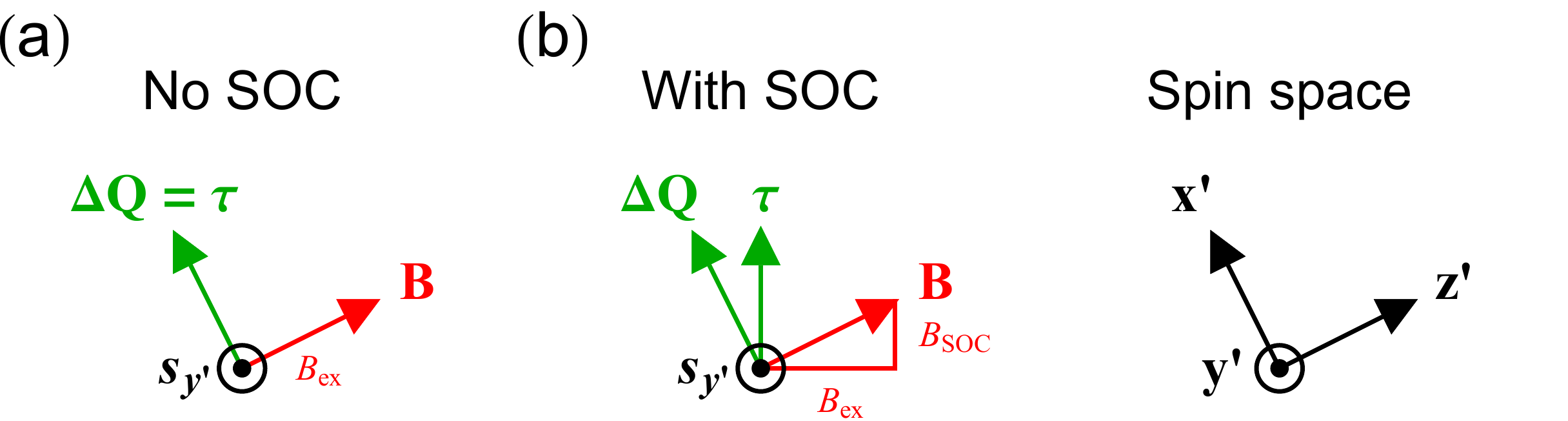}
	\caption{
	Result of adding spin-orbit coupling to the magnetic impurity. (a) Without spin-orbit coupling, the magnetic field of the impurity $\bvec{B}$ equals the exchange field $\bvec{B}_\text{ex}$. The absorbed spin current $\Delta Q$ equals the torque $\bvec{\tau} \propto \bvec{s} \times \bvec{B}_\text{ex}$. (b) With spin-orbit coupling, $\bvec{B}$ is the sum of the exchange field $\bvec{B}_\text{ex}$ and the spin-orbit field $\bvec{B}_\text{soc}$. The absorbed spin current is perpendicular to the total field, not the exchange field. However, only the part of the absorbed spin current that is perpendicular to the exchange field contributes to the torque on the magnetization.  The rest of the absorbed spin current exerts a torque on the lattice through the spin-orbit coupling.
	}
	\vspace{0pt}
	\label{AddSOC}
\end{figure}

We now consider the case in which the impurity's total magnetic field is given by $\bvec{B} = \bvec{B}_\text{ex} + \bvec{B}_\text{soc}$, where $\bvec{B}_\text{ex}$ denotes the exchange field while $\bvec{B}_\text{soc}$ is the effective magnetic field from the spin-orbit interaction. Let us assume again that the total magnetic field $\bvec{B}$ points along $\zhat'$ while $\bvec{B}_\text{ex}$ and $\bvec{B}_\text{soc}$ individually do not. As before, $\bvec{\tau}$ points along $x'$ and $\bvec{s}_0$ lies in the $y'/z'$ plane. The torque on the impurity's magnetic moment is due to the misalignment of the spin and the exchange field:\cite{haney2010current}
\begin{align}
    \bvec{\tau} \propto \bvec{s}_0 \times \bvec{B}_\text{ex}.
\end{align}
Since $\bvec{B}_\text{ex}$ is not required to point along $z'$ as before, there is no guarantee that the absorbed spin current equals the exchange torque, as can be seen in Fig.~\ref{AddSOC}). In other words, while the absorbed spin current must equal the torque on the total effective field $\bvec{B}$, it need not equal the torque on only part of that effective field (i.e. $\bvec{B}_\text{ex}$). In this case, spin-orbit coupling has introduced another channel of angular momentum transfer between the scattered electron and the lattice at the magnetic impurity.

Without spin-orbit coupling at the interface, the longitudinal spin current (spins aligned along the magnetization) is conserved, but the transverse spin current is not, as discussed in Sec.~\ref{subsec:SpinTorque}.  This no longer holds at interfaces with spin-orbit coupling.  It does hold for individual states with respect to the total effective field, but not with respect to the exchange field (aligned with magnetization) alone. This difference becomes important in Sec.~\ref{subsec:SpinAccumulation}.

\subsection{Spin currents and spin torques in a bilayer}

In the previous section, we showed that an electron transfers angular momentum to a magnetic impurity after scattering off of it. The change in spin flux always equals the total torque, but in the presence of spin-orbit coupling, the torque on the impurity's magnetic moment is only part of the total torque. Now, we consider a bilayer system in which the material interface plays the role of the magnetic impurity.

Our primary goal is to relate the accumulations and currents that drive transport with the resulting spin currents and spin torques at the interface. In equilibrium, we model both layers as free electron gasses with identical, spin-independent, spherical Fermi surfaces. We do so mainly for simplicity, but also to focus on the important qualitative behavior arising in nonequilibrium. To distinguish between layers, we assume each layer has different momentum relaxation times ($\tau$), and to model ferromagnetic layers, we assume $\tau$ is spin-dependent.

Our description of electron transport and the resulting spin currents and torques comes from the Boltzmann equation.  A full solution of the Boltzmann equation\cite{Xiao:2007} involves taking a general form for the solution of the bulk Boltzmann equation in each layer, computing matching conditions at their common interface, and applying boundary conditions based on the behavior of the solution at infinity.  While this approach is straightforward, it precludes simple analytical models for the results.  Here, we approximate the non-equilibrium distribution of electrons by neglecting the effect of scattering near the interface on the distribution functions, focusing on the matching conditions across the interface. This approach enables analytical solutions that are impossible if we consider the full solution. While these approximate solutions differ quantitatively from the full solution, they are qualitatively the same and allow full consideration of what processes are possible at interfaces.

At the interface, we adopt a basic quantum mechanical picture where electrons are plane waves with two spin degrees of freedom. Effective magnetic fields capture the exchange interaction and spin-orbit coupling at the interface, similar to the previous section. These effective magnetic fields behave like spin-dependent potential barriers, leading to spin-dependent scattering. From this scattering, spin currents and spin torques arise.

In the following, we formally define this model, describe the crucial approximations, and present the results without derivation, which can be found in Appendix~\ref{appendix:spintransport}.

\subsubsection{Boltzmann model with quantum mechanical interfacial scattering}

First, we formally define the model, beginning with a description of the interface. The effective magnetic field $\bvec{B}_\text{eff}$ seen by carriers at the interface is given by
\begin{align}
    \bvec{B}_\text{eff} &= \bvec{B}_\text{ex} + \bvec{B}_\text{soc} \propto \bvec{u}_\text{eff}
    \label{Beff}
\end{align}
where $\bvec{u}_\text{eff}$ is a unitless quantity proportional to the effective magnetic field (note that $\bvec{u}_\text{eff}$ is not a unit vector). Assuming Rashba-type spin-orbit coupling, $\bvec{u}_\text{eff}$ is given by:
\begin{align}
    \bvec{u}_\text{eff} &= u_\text{ex} \mhat + u_\text{R} \zhat \times \bvec{k}/k_F,
    \label{ueff}
\end{align}
where $\mhat$ is the unit vector pointing along the interfacial magnetization, $\bvec{k}$ is the crystal momentum of the electron traveling towards the interface, $k_F$ is the Fermi wave vector, and $u_\text{ex}$ and $u_\text{R}$ are unitless parameters describing the relative strengths of the exchange and spin-orbit interactions respectively. Throughout this section, we use the following parameterization:
\begin{align}
    u_\text{ex} = |\bvec{u}_\text{eff}| \cos(\uangle), \quad
    u_\text{R} = |\bvec{u}_\text{eff}| \sin(\uangle).
    \label{ExSOCparam}
\end{align}
This parameterization is useful because the results have a simple dependence on $\uangle$ (despite having a complicated dependence on $|\bvec{u}_\text{eff}|$). This simple $\uangle$ dependence allows us to probe the limits of vanishing exchange interaction or vanishing spin-orbit coupling.

Including a spin-independent potential barrier (parameterized by $u_0$), this system is described by the following $2\times2$ Hamiltonian
\begin{align}
    H(\rhat)    = \frac{\hbar^2 k^2}{2m} I_{2\times2}
                + \frac{\hbar^2 k_F}{m} \delta(z) \big{(}
                u_0 I_{2\times2} 
                + \bvec{\sigma} \cdot \bvec{u}_\text{eff}
                \big{)},
    \label{Vint}
\end{align}
where $I_{2\times2}$ is the identity matrix in spin space, and the interface is located at $z=0$. Note that the factor $\hbar^2 k_F \delta(z)/m$ converts the unitless vector $\bvec{u}_\text{eff}$ to units of energy. Wavefunction matching at the interface yields reflection and transmission amplitudes, which are derived in Appendix~\ref{appendix:spintorque}. To determine the resulting spin currents and spin torques, we use these reflection and transmission amplitudes as boundary conditions for the Boltzmann equation.

The statistics of carriers in each layer of the system are captured by the Boltzmann distribution function. In the model we consider here, with full coherence between all spin states at each point in reciprocal space but no coherence between different points in reciprocal space, the full distribution function can be captured by four functions, $f_\alpha(\bvec{r},\bvec{k})$ where $\alpha\in[x,y,z,c]$, representing the expectation value of spin along each axis and the number operator. 
Since we are interested in the linear transport regime, we can linearize the distribution function around its equilibrium form
\begin{align}
    f_\alpha(z,\bvec{k}) &= f_\text{eq}(\epsilon_{\bvec{k}}) \delta_{\alpha c} + \frac{\partial f_\text{eq}}{\partial \epsilon_{\bvec{k}}} g_\alpha(z,\bvec{k})
\end{align}
where $\epsilon_{\bvec{k}}$ is the $\bvec{k}$-dependent energy, $f_\text{eq}$ is the spin-independent equilibrium distribution function, $g_\alpha$ is the nonequilibrium perturbation of the distribution function, and $\alpha\in[x,y,z,c]$.  In the simple model for the electronic structure that we consider here, the equilibrium distribution is independent of spin, as reflected in the $\delta_{\alpha c}$ in the first term.

To evaluate the accumulations and currents at the interfaces, we must know the distribution functions $g_\alpha(z,\bvec{k})$ at $z = 0^\pm$. We could solve the Boltzmann equation for the entire bilayer using the quantum mechanical scattering matrices as boundary conditions. This process requires numerical solutions that are cumbersome to compare with experiments. In the next section, we outline a simple but effective approximation that bypasses a full solution of the Boltzmann equation for the bilayer.

\subsubsection{Obtaining qualitative results without solving the Boltzmann equation for the entire heterostructure}

\begin{figure*}[t!]
	\centering
	\vspace{0pt}	
	\includegraphics[width=0.95\linewidth,trim={0.0cm 1.5cm 0.0cm 2.5cm},clip]{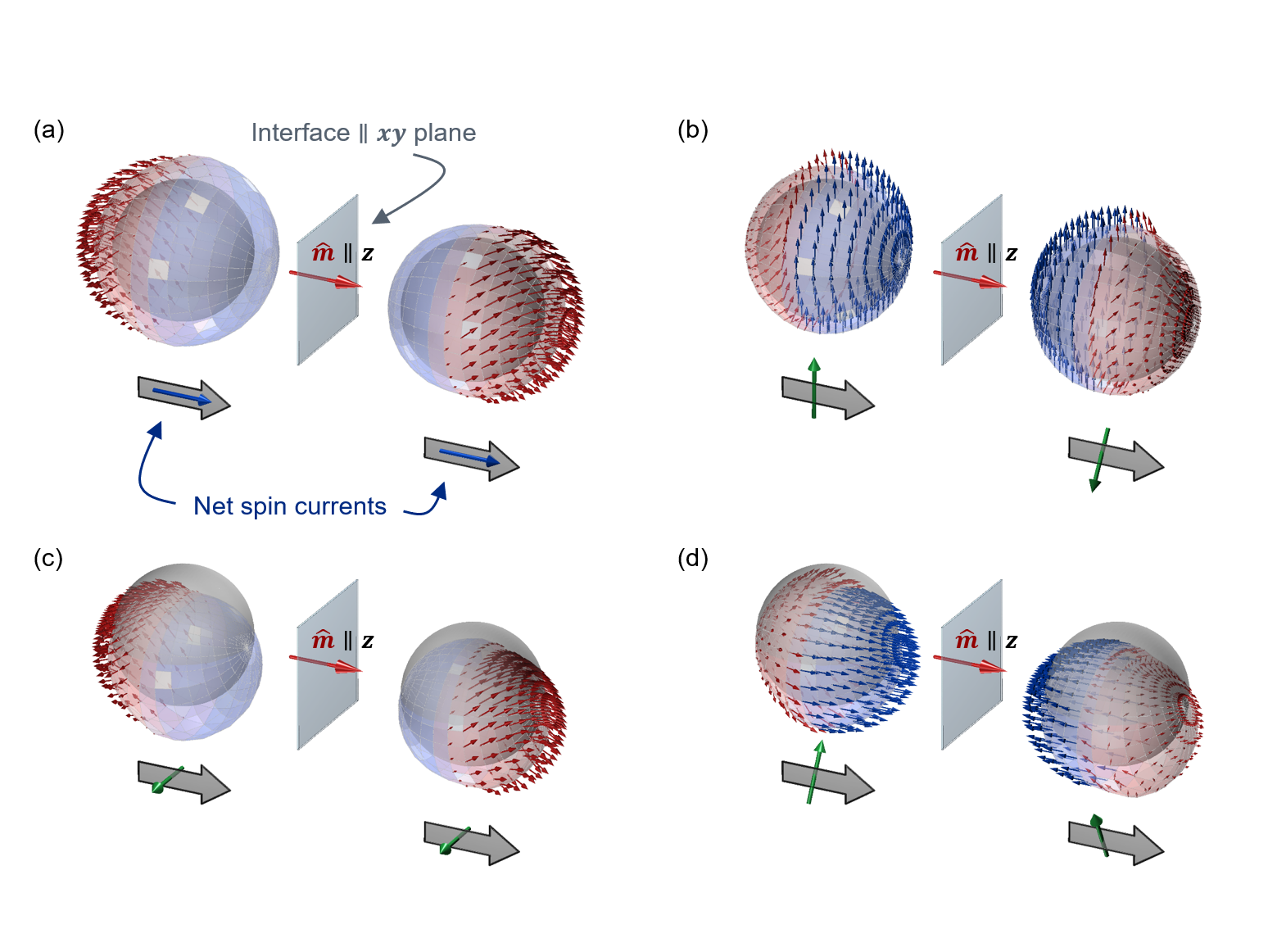}
	\caption{
	Plots depicting the nonequilibrium distribution functions $g_\alpha(0^\pm,\bvec{k})$ in the presence of interfacial spin-orbit scattering. Each picture illustrates the physics captured by a single matrix element in the matrices defined in \Eq{Aforz} or \Eq{Aforx}. The interfacial exchange field (red arrow) points out-of-plane (along $z$). The gray sphere represents the equilibrium Fermi surface. The colored surfaces represent the nonequilibrium perturbation to the Fermi surface, given by the charge distribution $g_c(0^\pm,\bvec{k})$. The arrows depict the spin distribution function $g_i(0^\pm,\bvec{k})$ for $i \in [x,y,z]$. Blue and red regions represent the incoming and outgoing (scattered) carriers respectively. The net spin current at $z = 0^\pm$ is shown below the distribution functions, where the block arrows denote spin flow (always out-of-plane) and the tubular arrows denote spin direction. Note that we use \emph{transverse} and \emph{longitudinal} to denote spin directions relative to the interfacial exchange field. (a) Scenario where the incident carriers have two different charge accumulations but no spin accumulation. Regardless, the scattered carriers are spin polarized from their interaction with the interfacial exchange and spin-orbit fields. The net spin currents after scattering have longitudinal spin directions and are conserved across the interface. (b) Scenario where the incident carriers have two different transverse spin accumulations. The net spin currents after scattering also have transverse spin directions but are rotated relative to the spin accumulation and not conserved across the interface. (c) Scenario where two different in-plane charge currents flow at $z = 0^\pm$, indicated by differing shifts the Fermi surface. The scattered carriers become spin polarized and the net out-of-plane spin currents have transverse spin direction. (d) Scenario where two different in-plane, longitudinal spin currents flow at $z = 0^\pm$. The net out-of-plane spin currents have transverse spin direction and are not conserved across the interface.
	}
	\vspace{0pt}
	\label{kDist}
\end{figure*}

To obtain a useful, quasi-analytical description of spin currents and spin torques at interfaces without solving the Boltzmann equation, we focus on two cases: 
\begin{enumerate}
    \item Perpendicular transport in which spin and charge accumulations at $z = 0^\pm$ drive out-of-plane currents
    \item In-plane transport that drives out-of-plane spin currents
\end{enumerate}
For each case, we assume a reasonable form for the distribution function of incoming electrons, defined by $k_z > 0$ for $z < 0$ and $k_z < 0$ for $z > 0$. The outgoing (scattered) electrons are then determined by the quantum mechanical scattering matrices. To simplify our notation, we rewrite the distribution function as a four-vector labeled by components $\alpha\in[x,y,z,c]$, such that $g_\alpha \rightarrow \fourvec{g}$. For perpendicular transport,  , we then approximate $\fourvec{g}$ for incoming carriers as
\begin{align}
    \fourvec{g}(0^\pm,\bvec{k}) = e \fourvec{\pu}_\pm ,
    \label{gacc}
\end{align}
where $\fourvec{\pu}_\pm$ is a constant four-vector defined separately on both sides of the interface $z=0^\pm$. Such distributions are shown in the blue regions of panels (a) and (b) of Fig.~\ref{kDist} for the cases of charge accumulation and spin accumulation respectively.  For in-plane electric fields, the incoming distribution is defined by 
\begin{align}
    \fourvec{g}(0^\pm,\bvec{k}) = e \tilde{k}_x \fourvec{\pu}_\pm ,
    \label{gip}
\end{align}
where the constant $\pu_{\alpha\pm}=E v_F \tau_{l\alpha} P_{\alpha\pm}$, $E$ is the magnitude of the in-plane electric field, $v_F$ the Fermi velocity, $\tau_{\alpha\pm}$ the momentum relaxation times, and $P_{\alpha\pm}$ the dimensionless spin or charge polarization with $\pm$ indicating either side of the interface. These distributions are shown in the blue regions of panels (c) and (d) of Fig.~\ref{kDist} for the cases of charge accumulation and spin accumulation respectively. The panels in Fig.~\ref{kDist} also give, in red, the outgoing distribution functions resulting from interfacial scattering for each of the different incoming distributions. When they scatter from the interface, incoming unpolarized carriers become spin polarized and incoming spin-polarized carriers rotate their spin polarization. These phenomena arise from the influence of the interfacial exchange and spin-orbit interactions, and lead to the modification of spin accumulations and spin currents at the interface, which we discuss in detail in the next subsection.

\subsubsection{Calculating the spin/charge accumulations and spin/charge currents at interfaces}

Using the incoming distributions as defined in \Eq{gacc} or \Eq{gip}, we use the matching conditions at the interface (discussed below) to determine the full distributions shown in Fig.~\ref{kDist}. 
From these distribution, we can compute the non-equilibrium spin and charge densities (also called spin and charge accumulations) and the non-equilibrium spin and charge currents. At the two sides of the interface, $z = 0^\pm$, these quantities are given by the following integrals over the spherical Fermi surfaces
\begin{align}
    \fourvec{\mu}(0^\pm) &= c_\mu \int_\text{FS} d^2k \fourvec{g}(0^\pm,\bvec{k})     \label{s0def} \\
    \fourvec{j}_{z}(0^\pm) &= c_j \int_\text{FS} d^2k k_z \fourvec{g}(0^\pm,\bvec{k}),
    \label{scdef}
\end{align}
where $c_s$ and $c_j$ are constants. The components of the four-vector $\fourvec{\mu}$ represent the spin ($\alpha = x,y,z$) and charge ($\alpha=c$) accumulations at $z=0$. The components of $\fourvec{j}_z$ similarly represent the spin and charge current flowing out-of-plane at $z=0^\pm$.  The conductance matrix would be constructed by solving for $\pu_\pm$ in terms of the accumulations $\fourvec{\mu}(0^\pm)$ and inserting those into the expression for $\fourvec{j}_{z}(0^\pm)$, see  Refs.~\onlinecite{iSOCAminFormalism,iSOCAminPhenomenology}.  Here, we do not reproduce the derivation of that matrix but focus on the different physical processes that contribute to it.

The dimensionless vector $\bvec{u}_\text{eff}$ characterizing the potential given in \Eq{ExSOCparam} describes the interaction between the exchange potential and the spin-orbit potential.  If we choose the magnetization to be out-of-plane, $\mhat = \zhat$, the exchange field and the spin-orbit field are always perpendicular to each other, greatly simplifying the form of the results. We present results for this case because intermediate results can be cast in a physically transparent form that allows for a clear understanding of the role played by the exchange interaction and spin-orbit coupling in the processes that occur at the interface.  The final result, obtained after integrating over the full Fermi surface obscures these simple roles. For in-plane or general direction magnetizations, the physics is the same but even the intermediate forms are complicated enough to obscure the physical interpretation. The full results can be found numerically as is done in Refs.~\onlinecite{iSOCAminFormalism,iSOCAminPhenomenology}. 

For a magnetization $\mhat = \zhat$ the vector $\bvec{u}_\text{eff}$ depends on each electrons' wave vector in two ways, most easily seen in spherical coordinates
\begin{align}
k_x &= k_F \sin(\theta)\cos(\phi) \\
k_y &= k_F \sin(\theta)\sin(\phi) \\
k_z &= k_F \cos(\theta).
\end{align}
The relative strength of the spin-orbit interaction depends on the polar angle $\theta$, going to zero as $\theta$ goes to zero and its direction depends on the azimuthal angle $\phi$. It turns out we can analytically evaluate the integrals in \Eqs{s0def}{scdef} over azimuthal angle $\phi$ using the definition for $\fourvec{g}$ given by \Eq{gacc} or \Eq{gip}. We refrain from evaluating the remaining integral over polar angle $\theta$ because it is cumbersome and not necessary to obtain physical insight. Even though the azimuthal average of the spin-orbit potential is zero, it still makes substantial contributions to the transport when the average is weighted by the distribution functions with either a spin-dependence or an angular dependence.  Carrying out these azimuthal integrations highlights the effects that remain.

In the following, we write the results in terms of the average and difference in values of $\fourvec{j}_z$ and $\fourvec{\pu}$ across the interface:
\begin{align}
    \Delta\fourvec{j}_{z} &= \frac{1}{2} \big{(}\fourvec{j}_{z}(0^-) - \fourvec{j}_{z}(0^+) \big{)} \quad
    \Delta\fourvec{\pu} = \frac{1}{2} (\fourvec{\pu}_- - \fourvec{\pu}_+) \\
    \bar{\fourvec{j}}_{z} &= \frac{1}{2} \big{(} \fourvec{j}_{z}(0^-) + \fourvec{j}_{z}(0^+) \big{)} \quad
    \bar{\fourvec{\pu}} = \frac{1}{2} (\fourvec{\pu}_- + \fourvec{\pu}_+)
    \label{DiffAvgdef}
\end{align}
Using this notation, the accumulations and out-of-plane currents are given by
\begin{align}
    \fourvec{\mu} &= \int_0^{\pi/2} d\theta w(\theta)
   \SMt(\theta) \bar{\fourvec{\pu}} \label{Tpdef} \\
    \bar{\fourvec{j}}_{z} &= \int_0^{\pi/2} d\theta v(\theta) 
    \SMp(\theta)  \Delta\fourvec{\pu}, \label{Sdef} \\
    \Delta\fourvec{j}_{z} &= \int_0^{\pi/2} d\theta v(\theta) 
    \SMm(\theta) \bar{\fourvec{\pu}},
    \label{Adef}
\end{align}
where $w(\theta) = c_u \tan(\theta)$, $v(\theta) = c_j e k_F^2 \sin(\theta)$, and $\SMt$, $\SMp$, and $\SMm$ are $4\times4$ matrices. The constants $c_\mu$ and $c_j$ are defined in the appendix. These quantities capture the matching conditions for the distribution functions at the interface based on transmission and reflection probabilities and the form of the incident distribution, accumulations versus in-plane electric field.  The magnitudes of the incident distributions are contained in $\fourvec{\pu}_\pm$  The index $\irtl=\{ \irl, \tl\}$ indicates whether the matrix refers to the \textit{incident plus reflected} side or the \textit{transmitted} side. Since we have made the approximation that the electronic structure is the same on both sides of the interface, the transmission and reflection probabilities are the same for electrons incident from the right and from the left.  These matrices are given by the integration over azimuthal angle of the appropriate transmission and reflection coefficients.  They are related as follows:
\begin{align}
    \SMp(\theta) &= \SMir(\theta) + \SMt(\theta)  \label{SMatdef} \\
    \SMm(\theta) &= \SMir(\theta) - \SMt(\theta). \label{AMatdef}
\end{align}
Note that the spin and charge accumulations defined in \Eq{Tpdef} are at the interface and differ from those, defined in \Eq{scdef}, on either side of the interface. 

Equations~\ref{Tpdef}-\ref{AMatdef} relate the important physical quantities in terms of the boundary conditions. They show that the symmetric response matrix $\SMp$ determines the average spin and charge currents at the interface ($\bar{\fourvec{j}}_z$) while the antisymmetric response matrix $\SMm$ determines the difference in spin and charge currents across the interface ($\Delta \fourvec{j}_z$). The form of $w$, $v$, $\SMir$, and $\SMt$ depends greatly on choice of $\fourvec{g}$, i.e whether accumulations or in-plane currents drive the system. In the next two subsections, we present the $\SMir$, and $\SMt$ matrices, and discuss how they capture the effect of interfacial spin-orbit coupling for both scenarios.

\subsubsection{Spin or charge accumulations drive the system}
\label{subsec:SpinAccumulation}

The choice of $\fourvec{g}(0^\pm,\bvec{k}) = -e\fourvec{\pu}_\pm$ corresponds to a spin and/or charge accumulation at $z = 0^\pm$ as might be driven by a perpendicular voltage or by the spin Hall effect for in-plane transport. In this case, the $\SMir$, and $\SMt$ matrices are:
\begin{widetext}
\begin{align}
    \SMirt =
    \begin{pmatrix}
    0           	&	0	            &	0               &	0	\\
    0	        	&	0           	&	0               &	0	\\
    0           	&	0	            &	0               &	0	\\
    0	        	&	0            	&	0               &	2c
    \end{pmatrix}
    + 2\fc
    \begin{pmatrix}
    \fc a_\irtl     &	-b_\irtl        &	0               &	0	\\
    b_\irtl        	&	\fc a_\irtl     &	0               &	0	\\
    0           	&	0	            &	\fc c           &	d	\\
    0	        	&	0            	&	d               &	0
    \end{pmatrix}
    + \fs^2
    \begin{pmatrix}
    a_\irtl + c   	&	0	            &	0               &	0	\\
    0	        	&	a_\irtl + c     &  	0               &	0	\\
    0           	&	0	            &	2a_\irtl        &	0	\\
    0	        	&	0            	&	0               &	0
    \end{pmatrix}.
    \label{Aforz}
\end{align}
\end{widetext}
where $\irtl \in [\irl,\tl]$. The parameters $a_\irtl$, $b_\irtl$, $c$, and $d$ depend only on the polar angle $\theta$, the magnitude of the effective field $|\bvec{u}_\text{eff}|$, and the spin-independent barrier strength $u_0$.  As a reminder, the indices in order are $[x,y,z,c]$. The importance of the spin-orbit interaction to the scattering depends on the wave vector.  For normal incidence it is zero and is maximal for grazing incidence.  Recall that the angle $\uangle$ defined in \Eq{ExSOCparam} reflects the relative importance of the exchange interaction and the spin-orbit coupling. For a particular angle of incidence $\theta$, the dependence of $\SMirt$ on angle $\uangle$ is given by the functions $\fs$ and $\fc$:
\begin{align}
    \fs &=  \frac{\sin(\uangle)}{\sqrt{\sin^2(\theta)\cos^2(\uangle)+\sin^2(\uangle)}} \\
    \fc &=  \frac{\sin(\theta) \cos(\uangle)}{\sqrt{\sin^2(\theta)\cos^2(\uangle)+\sin^2(\uangle)}},
    \label{SCI}
\end{align}
These obey the following limits:
\begin{align}
    \fsf(0) &= 0,
    \quad \fsf(\pi/2) = 1, \\
    \fcf(0) &= 1,
    \quad \fcf(\pi/2) = 0,
\end{align}
Therefore, in the limit of vanishing spin-orbit coupling ($\uangle = 0$) or vanishing exchange interaction ($\uangle = \pi/2$), only one of these functions is nonzero. The parameters $c$ and $d$ derive from the components of the spin longitudinal to the effective field, so they are conserved across the interface and hence are equal for the incident plus reflected and transmitted sides. They do enter some of the coefficients for the components transverse to the magnetization, because for each incident wave vector, the effective field is not along the magnetization. For this special case with the magnetization always perpendicular to the spin-orbit coupling field, all but a few of the possible contributions of this type get integrated away.  The parameters  $a_\irtl$, $b_\irtl$ are different on the incident plus reflected and transmitted sides as they are associated with the components of the spin current with spins transverse to the effective field. For this particular magnetic field, they only make a limited but important contribution to the transport for spins longitudinal to the field.

\begin{figure*}[]
	\centering
	\vspace{0pt}	
	\includegraphics[width=0.9\linewidth,trim={0.0cm 0.0cm 0.0cm 0.0cm},clip]{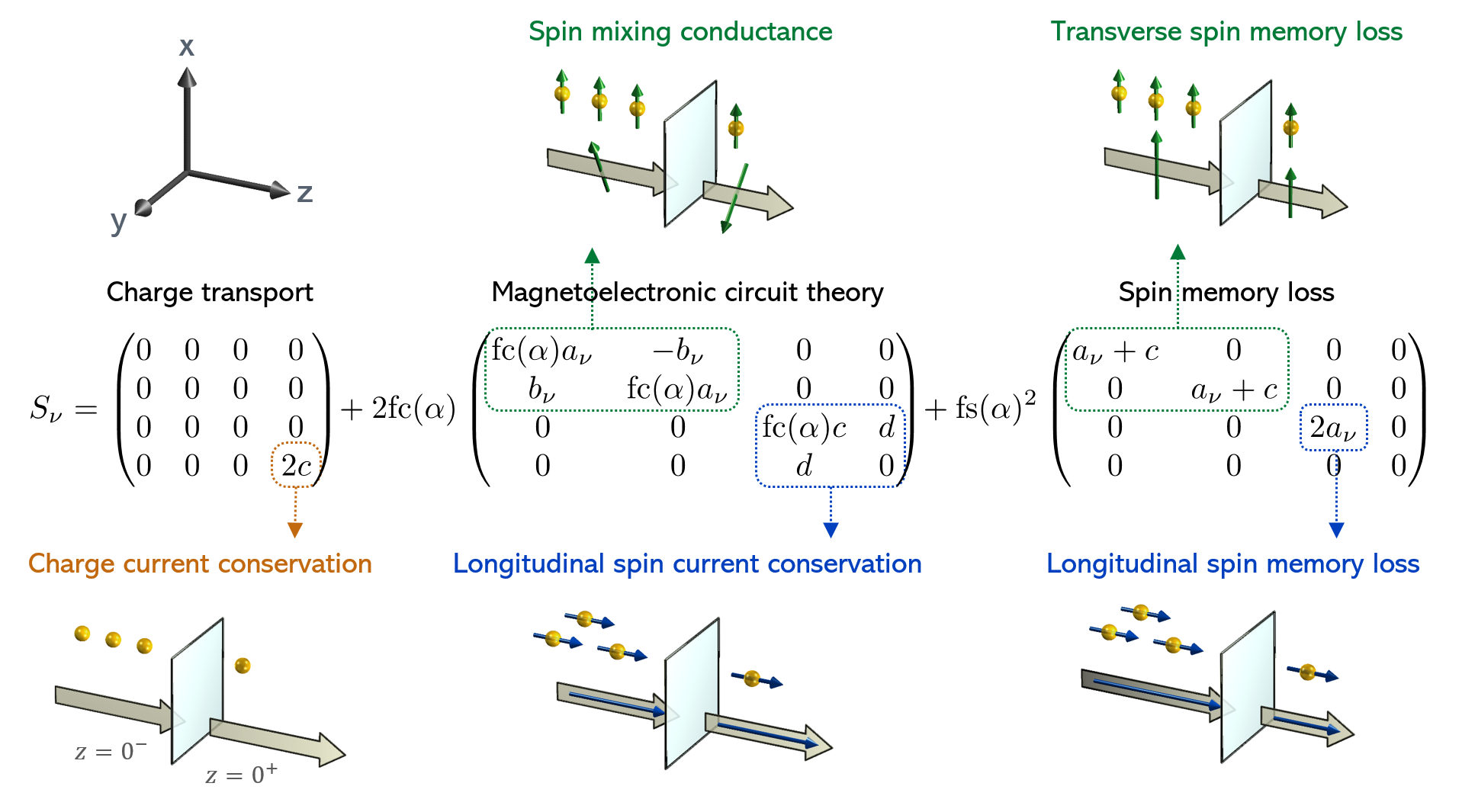}
	\caption{
	Breakdown of the $\SMirt$ matrices ($\irtl \in [\irl,\tl]$) when spin or charge accumulations drive transport at interfaces. The matrix $\SMt$ determines the spin and charge accumulation $\fourvec{\mu}$ at the interface (see \Eq{Tpdef}). The symmetric response $\SMp = \SMir + \SMt$ determines the average spin current $\bar{\fourvec{j}}_z$ at the interface (see \Eq{Sdef}). The antisymmetric response $\SMm = \SMir - \SMt$ determines the difference in spin current $\Delta\fourvec{j}_z$ across the interface (see \Eq{Adef}). The matrix column specifies the spin and charge accumulations at $z = 0^\pm$ while the row gives the components of $\fourvec{\mu}$, $\bar{\fourvec{j}}_z$, or $\Delta\fourvec{j}_z$, depending on whether \Eq{Tpdef}, \Eq{Sdef}, or \Eq{Adef} is used. The images depict the charge accumulations (gold spheres) or the spin accumulations (gold spheres with arrows) that drive the system and the resulting spin currents at $z = 0^\pm$, where block arrows denote flow direction and tubular arrows denote spin direction. 
	}
	\vspace{0pt}
	\label{AmatZ}
\end{figure*}

Eq.~\ref{Aforz} shows how spin/charge accumulations drive transport and clarifies the role of the exchange and spin-orbit interactions. Since $\mhat=\zhat$, we refer to spin polarizations along $x$ and $y$ as \emph{transverse} and those along $z$ as \emph{longitudinal}. Fig.~\ref{AmatZ} illustrates the contributions from each element of the $\SMir$ and $\SMt$ matrices and can be used as a companion to the main text below.

\emph{Charge current conservation}---The first matrix in \Eq{Aforz} describes charge current conservation across the interface. The sole nonzero parameter $2c$ relates the drop in charge accumulation across the interface to the total charge current flowing across the interface.

\emph{Generalized Magnetoelectronic Circuit Theory}---The second matrix in \Eq{Aforz} describes a generalization of magnetoelectronic circuit theory.\cite{MCTBrataas,MCTBrataas2} Because this matrix is multiplied by $2\fc$, it vanishes for zero exchange interaction ($\uangle = 0$). For a nonmagnet/ferromagnet bilayer, the real and imaginary parts of the spin mixing conductance are given by integrating $a_\irl$ and $b_\irl$ using \Eq{SCI}. The mixing conductance is also generalized to include a \emph{transmitted spin mixing conductance} given by integrating $a_\tl$ and $b_\tl$ using \Eq{SCI}. The concept of a transmitted mixing conductance has been discussed before\cite{brataas2003magnetoelectronic} and describes the part of the transverse spin current transmitted through the interface. The factor $\fc$ generalizes the real part of the mixing conductance to capture features of interfacial spin-orbit coupling.

The top left $2\times2$ block relates the transverse spin accumulations at $z = 0^\pm$ to the transverse spin currents at $z = 0^\pm$ and the transverse spin accumulation at $z = 0$. The transverse spin currents are not conserved across the interface since $a_\irl \neq a_\tl$ and $b_\irl \neq b_\tl$. For vanishing spin-orbit coupling, the transverse spin current at $z=0^-$ gives the total spin transfer torque. 

The bottom right $2\times2$ block relates the longitudinal spin accumulation and charge accumulation at $z = 0^\pm$ to the longitudinal spin current and charge current at $z = 0^\pm$. Because this block depends only on $c$ and $d$, the longitudinal spin currents governed by this block are conserved across the interface.

\emph{Spin Memory Loss}---Finally, the third matrix in \Eq{Aforz} captures spin memory loss. Because this matrix is multiplied by $\fs^2$, it vanishes for zero spin-orbit coupling ($\alpha=\pi/2$). The three nonzero matrix elements parameterize spin memory loss, which we describe as a magnitude difference in the spin current driven by spin/charge accumulations at $z = 0^\pm$. While all spin components experience spin memory loss (since $a_\irl \neq a_\tl$), the degree of spin memory loss differs for transverse and longitudinal spin currents.

When the magnetization is oriented in any other direction than normal to the interface, the form of these results becomes much more complicated. The four by four matrix loses the diagonal two by two simplification. However, the same processes described above still take place, though their effects change quantitatively and get spread out throughout the matrix.

\subsubsection{In-plane spin or charge currents drive the system}

\begin{figure*}[]
	\centering
	\vspace{0pt}	
	\includegraphics[width=0.9\linewidth,trim={0.0cm 0.0cm 0.0cm 0.0cm},clip]{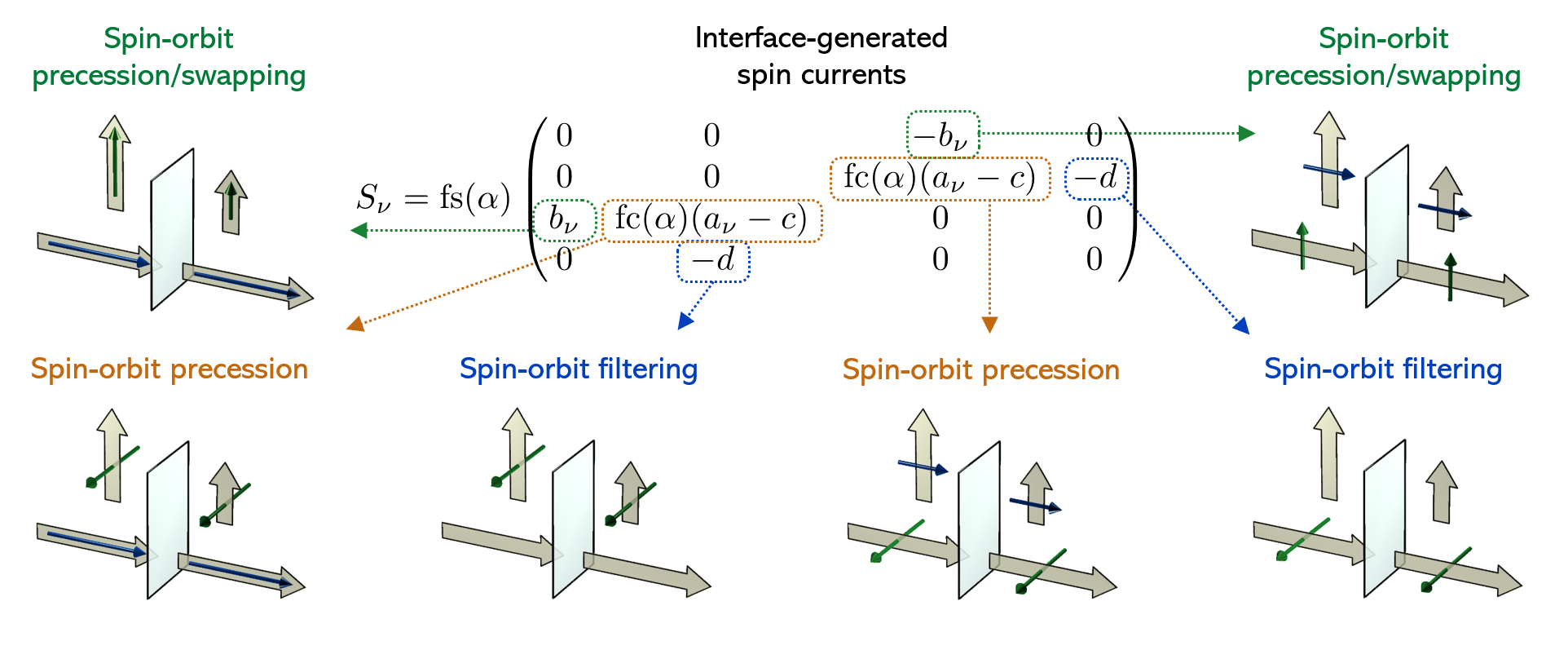}
	\caption{
	Breakdown of the $\SMirt$ matrices when in-plane spin/charge currents drive transport at interfaces. As in Fig.~\ref{AmatZ}, the matrix $\SMt$ determines the spin/charge accumulation $\fourvec{\mu}$ at the interface (see \Eq{Tpdef}), the symmetric response $\SMp = \SMir + \SMt$ determines the average spin current $\bar{\fourvec{j}}_z$ at the interface (see \Eq{Sdef}), and the antisymmetric response $\SMm = \SMir - \SMt$ determines the difference in spin current $\Delta\fourvec{j}_z$ across the interface (see \Eq{Adef}). The column specifies the in-plane spin/charge currents at $z = 0^\pm$ while the row gives the components of $\fourvec{\mu}$, $\bar{\fourvec{j}}_z$, or $\Delta\fourvec{j}_z$, depending on whether \Eq{Tpdef}, \Eq{Sdef}, or \Eq{Adef} is used.  The images depict both the in-plane and out-of-plane spin currents at $z = 0^\pm$ using block arrows for flow direction and tubular arrows for spin direction.
	}
	\vspace{0pt}
	\label{AmatX}
\end{figure*}

The choice of $\fourvec{g} = \tilde{k_x} \fourvec{\pu}_\pm$ corresponds to an in-plane driving current at $z = 0^\pm$ (here flowing along $x$). The components $\pu_x$, $\pu_y$, and $\pu_z$ describe the spin polarization of the $x$-flowing spin current driving the system while $\pu_c$ describes the $x$-flowing charge current driving the system. For this system, the $\SMirt$ matrices are given by:
\begin{align}
    \SMirt = 
    \fs
    \begin{pmatrix}
    0		&	0	            &	-b_\irtl        &	0	\\
    0	  	&	0	           	&	\fc(a_\irtl-c)  &   -d	\\
    b_\irtl	&	\fc(a_\irtl-c)  &	0	            &   0	\\
    0	   	&	-d              &	0	            &	0	
    \end{pmatrix}.
    \label{Aforx}
\end{align}
The parameters $a_\irtl$, $b_\irtl$, $c$, and $d$ are the same as those used in Eq.~\ref{Aforz}. Since $\SMirt$ is proportional to $\fs$, we see that a nonzero interfacial spin-orbit interaction is required to couple an in-plane driving current with out-of-plane spin currents or a spin torque.  As a reminder, the indices in order are $[x,y,z,c]$. 
Below, we discuss the important features of \Eq{Aforx}, which are also illustrated in Fig. \ref{AmatX}.

\emph{Generalized Rashba-Edelstein effect}---First, we show that in-plane currents create interfacial spin accumulations. The spin/charge accumulations at $z = 0$ are given by
\begin{align}
    \fourvec{\mu}(0) &= \int_0^{\pi/2} d\theta w(\theta) \SMt(\theta) \bar{\fourvec{\pu}}
\end{align}
and thus governed by the $\SMt$ matrix. As seen from \Eq{Aforx}, the parameter $-d$ in the second row, fourth column of $\SMt$ relates an in-plane charge current to the spin accumulation along $y$. This describes the Rashba-Edelstein effect. Spin accumulations in other directions only arise when in-plane \emph{spin currents} drive the system. In ferromagnet/nonmagnet bilayers, an in-plane charge current becomes spin polarized along the magnetization $\mhat$ in the ferromagnetic layer (here $\mhat = \zhat$). According to the first row, third column in \Eq{Aforx}, an in-plane spin current polarized along $z$ creates a spin accumulation along $x$. To describe these additional spin accumulations arising from in-plane spin currents, we use the term \emph{generalized Rashba-Edelstein effect}.

\emph{Spin-orbit filtering}---Second, we show that an in-plane charge current (here along $x$) generates an out-of-plane spin current (here along $z$) with spin direction along $y$. This spin current shares the same orientation as the spin Hall current. According to our model, when in-plane currents differ at $z = 0^-$ and $z = 0^+$, an out-of-plane spin/charge current develops across the interface. This out-of-plane current is given by
\begin{align}
\bar{\fourvec{j}}_{z} &= \int_0^{\pi/2} d\theta v(\theta) \SMp(\theta) \Delta\fourvec{\pu}
\end{align}
where $\SMp = \SMir + \SMt$. According to \Eq{Aforx}, in-plane charge currents create out-of-plane spin currents with spin direction along $y$. Although not strictly an inverse effect, \Eq{Aforx} suggests that in-plane spin currents with spin direction $y$ also result in out-of-plane charge currents. Both of these effects are proportional to $d$. We call both effects \emph{spin-orbit filtering}, because they result from electron spins being filtered by the spin-orbit field while scattering off the interface.

\emph{Spin-orbit precession}---Finally, we show that in-plane spin currents generate out-of-plane spin currents at interfaces. To see this, note that the parameter $b_\irtl$ (which appears twice in \Eq{Aforx}) describes the following two cases: 1) an $x$-flowing spin current with $z$-spin direction creates a $z$-flowing spin current with $x$-spin direction and 2) an $x$-flowing spin current with $x$-spin direction creates a $z$-flowing spin current with $z$-spin direction. Both of these cases are phenomenologically identical to \emph{spin swapping}, where nonmagnets convert spin currents into other spin currents by swapping their flow and spin directions. However, the terms in \Eq{Aforx} proportional to $\fc(a_\irtl-c)$ do not follow the spin swapping mechanism, but nevertheless convert in-plane spin currents into out-of-plane spin currents.  To unify these concepts, we refer to this family of effects at interfaces as \emph{spin-orbit precession}, because they result from electron spins rotating about the spin-orbit field while scattering off the interface.

The spin currents generated at interfaces are not necessarily identical at $z = 0^-$ and $z = 0^+$. This discontinuity in spin current across the interface is given by
\begin{align}
\Delta\fourvec{j}_{z} &= \int_0^{\pi/2} d\theta v(\theta) \SMm(\theta) \bar{\fourvec{\pu}},
\end{align}
where nonvanishing terms in the antisymmetric response $\SMm = \SMir - \SMt$ contribute to the discontinuities. Inspection of \Eq{Aforx} reveals that spin-orbit precession currents are discontinuous at the interface. In general, both spin-orbit filtering and spin-orbit precession currents can be discontinuous at the interface. Here, the continuity of spin-orbit filtering currents is a result of the simplicity of this model.

Together, Eq.~\ref{Aforz} and Eq.~\ref{Aforx} capture the processes that contribute to spin-orbit torques when the magnetization is perpendicular to the interface. Eq.~\ref{Aforx} captures the direct processes due to an in-plane electric field at the interface between two different materials and Eq.~\ref{Aforz} captures the processes that are initiated in the interior of the layers through effects like the spin Hall effect that gives rise to a spin current scattering from the interface. The relevant parts of the incoming distribution functions are combined with the relevant $\SMirt$ matrices to give the interfacial torques through the interfacial spin accumulation. The same matrices give the outgoing spin currents. Those directed into the ferromagnet typically dephase and contribute to the torque on that layer.  Those directed into the non-magnetic layer can traverse that layer and in trilayers contribute to the torque on the other ferromagnetic layer.

\section{Outlook}

In the previous section, we introduced a quasi-analytical model that captures how spin-orbit scattering at interfaces generates out-of-plane spin and charge currents and spin torques. These currents and torques were studied for two driving mechanisms: 1) spin/charge accumulations form on each side of the interface and 2) in-plane spin/charge currents flow on each side of the interface. The system could be nonmagnetic or contain a ferromagnetic layer. In the latter case, magnetism at the interface came from an interfacial exchange interaction while magnetism in the bulk layers was omitted in the electronic structure; however, the spin-polarized current in the ferromagnetic layer was captured via spin-dependent momentum relaxation times. When in-plane spin currents drive the system, we allow their spin direction to be longitudinal or transverse to the ferromagnetic layer's magnetization, capturing symmetry-allowed spin currents that are typically not considered in such systems.

Although the bulk layers in the model have a trivial electronic structure, the driving mechanisms we consider are fairly general, allowing exploration of many scenarios, albeit qualitatively. For instance, in nonmagnet/ferromagnet bilayers, the spin Hall effect generates a spin accumulation at the interface which exerts a torque on the ferromagnetic layer. The spin Hall effect arises from an in-plane charge current, and we find that this in-plane charge current also generates an out-of-plane spin current at the interface. This interface-generated spin current can have a different spin direction than the spin Hall current, thus enabling different torques. 

The model also describes the role of in-plane spin currents in the ferromagnetic layer when generating spin currents and spin torques. For example, in-plane charge currents are spin-polarized in ferromagnets along the magnetization direction. Near the interface, the electrons carrying this spin-polarized current interact with interfacial spin-orbit fields, which rotate their spin polarization and generate spin accumulations not captured by the two-dimensional inverse galvanic effect (or Rashba-Edelstein effect). We also consider in-plane spin currents with spin direction \emph{transverse} to the magnetization, which are allowed by symmetry but not well studied in the context of spin-orbit torque. These in-plane spin currents generate out-of-plane spin currents that also exert torques not predicted by traditional models that omit three-dimensional spin-orbit scattering. We show that this family of effects, which we called \emph{spin-orbit precession}, includes phenomena like spin swapping that was first predicted in nonmagnets\cite{SSTheoryLifshitsDyakonov} and later studied in ferromagnetic systems.\cite{SSTheorySaidaoui,SSTheoryPauyac}

Moving away from bilayers, we can also consider spin currents created in other layers not adjacent to the interface, as in spin valves. Such spin currents can eventually flow across the interface and undergo spin memory loss. If one of the layers adjacent to the interface is ferromagnetic, the degree of spin memory loss differs for spin currents with transverse and longitudinal spin directions (where transverse and longitudinal are defined relative to the magnetization).

The phenomena discussed here only scratch the surface of what is allowed at interfaces with spin-orbit coupling. Various magnetoresistance effects (like the spin Hall magnetoresistance) should be affected by spin-orbit scattering at interfaces. Following the methods in this paper, one may extend our model to describe how in-plane electric fields generate in-plane spin and charge currents near interfaces that are modulated by magnetization direction. Thus, simple extensions to this model should capture the effect of interfacial spin-orbit scattering on current-in-plane magnetoresistance effects.

Experiments have yet to verify many of these theoretical predictions. Part of the difficulty comes from the lack of reliable experimental techniques to independently quantify bulk and interfacial contributions to spin torques. We do not offer a solution to this problem. However, some of the difficulty also arises from bilayer systems, where the sum of several effects are lumped into a single measurement. Experiments in ferromagnetic multilayers have already shown the existence of competing torques that each damp the magnetization towards two separate axes;\cite{SOTExpHumphries,iSOCBaekAmin,iSOCHibino} this phenomena could be explained by the spin-orbit precession effects discussed earlier. By giving a clear, qualitative picture of what interfacial spin-orbit scattering enables, we hope to guide new experiments that can probe these effects (perhaps in unconventional heterostructures), and motivate new methods to electrically control magnetization dynamics.

\section*{Acknowledgments}
Work by V.P.A. was supported by Quantum Materials for Energy Efficient Neuromorphic Computing, an Energy Frontier Research Center funded by the U.S. Department of Energy (DOE), Office of Science, Basic Energy Sciences (BES), under Award \#DE-SC0019273. The authors appreciate useful comments from Robert McMichael, Jabez McClelland, Hans Nembach, Ivan Schuller, Andrew Kent, and Axel Hoffmann.

\appendix

\section{Spin torques in bilayers}
\label{appendix:spintorque}

First, we derive the quantum mechanical scattering amplitudes relevant to the phenomenological model. In this model, only the interface between layers has magnetism, which is captured by an effective magnetic field $\bvec{B}$. A free electron gas describes the bulk of each layer while a delta function potential describes the interface. Although this model is three-dimensional, it reduces to the one-dimensional model derived earlier for each incoming electron, except now we relax the condition that $\uU \rightarrow \infty$ and $\uD = 0$.

The $2\times2$ Hamiltonian for the system is,
\begin{align}
    H(\rhat)    &= \frac{\hbar^2 k^2}{2m} I_{2\times2} 
                + \delta(z) \big{(} V_0 I_{2\times2} 
                + J_\text{ex} \bvec{\sigma} \cdot \hat{\bvec{B}} \big{)}
    \label{H}
\end{align}
where the spin-independent potential $V_0$ and interfacial exchange energy $J_\text{ex}$ can be written as: 
\begin{align}
V_0 = \hbar^2 k_F (u_\UA + u_\DA) / 2m \\
J_\text{ex} = \hbar^2 k_F (u_\UA - u_\DA) / 2m
\end{align}
Here $k_\text{F}$ is the Fermi momentum (which is the same for both layers) and $\uUD$ is the unitless spin-dependent barrier strength at the interface. 

Alternatively, we can write this Hamiltonian explicitly in the spin basis aligned with the effective magnetic field $\bvec{B}$:
\begin{align}
    H(\rhat)    &= \frac{\hbar^2}{m} \begin{pmatrix} 
                k^2/2 + \delta(z) k_F u_\UA & 0 \\ 
                0 & k^2/2 + \delta(z) k_F u_\DA \end{pmatrix}
    \label{Hmat}
\end{align}
In this form, the problem reduces to two independent channels for spins parallel or antiparallel with $\bvec{B}$.


Consider an electron scattering off the interface. The electron arrives at the interface in one layer (layer 1) and is either reflected back into this layer or transmitted into the other layer (layer 2). Assuming that during scattering, the electron's in-plane momentum is conserved (specular scattering), the wavefunctions in layers 1 and 2 are given by
\begin{align}
    \psi_1(\bvec{r})   &= e^{\im \bvec{k}_\perp \cdot \bvec{r}_\perp} \big{(}
                \chi_\inc e^{\im k_z z} + \chi_\refl e^{-\im k_z z} \big{)}, \\
    \psi_2(\bvec{r})   &= e^{\im \bvec{k}_\perp \cdot \bvec{r}_\perp} \chi_\trans e^{\im k_z z},
    \label{WF}
\end{align}
where $z$ is the out-of-plane direction, $k_z$ is the out-of-plane component of momentum, and $\bvec{r}_\perp$ and $\bvec{k}_\perp$ are the in-plane position and momentum vectors, such that $\bvec{k} = (\bvec{k}_\perp, k_z)$ and $\bvec{r} = (\bvec{r}_\perp, z)$. The spinors $\chi_\inc$, $\chi_\refl$, and $\chi_\trans$ describe the incoming, reflected, and transmitted states respectively.

The reflected and transmitted wavefunctions are related to the incoming wavefunction through the scattering matrices. Thus, we may assume, for some $2\times2$ matrices $r$ and $t$ that $\chi_\refl = r \chi_\inc$ and $\chi_\trans = t \chi_\inc$. Assuming the interface lies at $z = 0$, we have:
\begin{align}
 \left.\begin{aligned}
    \psi_1              &= (1 + r) \chi_\inc, \\
    \partial_z \psi_1   &= \im k_z (1 - r) \chi_\inc, ~
       \end{aligned}
 \right \}
 ~~ z = 0^- 
  \label{WF1} \\
  \left.\begin{aligned}
    \psi_2              &= t \chi_\inc, \\
    \partial_z \psi_2   &= \im k_z t \chi_\inc, ~
       \end{aligned}
 \right \}
 ~~ z = 0^+
 \label{WF2}
\end{align}
Due to in-plane momentum conservation, the scattering problem has now been reduced to a one-dimensional problem defined along $z$.

Boundary conditions dictate that the wavefunction and particle current match at $z = 0^-$ and $z = 0^+$, which gives:
\begin{align}
    1 + r &= t, \label{BC1} \\
    1 - r^\dagger r &= t^\dagger t. \label{BC2}
\end{align}
The latter condition arises from matching the probability current
\begin{align}
    j   &= \frac{\hbar}{2m\im} \Big{(} \psi^\dagger (\partial_z \psi) 
        - (\partial_z \psi^\dagger) \psi \Big{)},
    \label{PC}
\end{align}
at $z = 0^-$ and $z = 0^+$, and can be checked using \Eqs{WF1}{WF2}. The spin density ($s_i$) and out-of-plane flowing spin current ($Q_{zi}$) are
\begin{align}
    s_{i}   &= \psi^\dagger \sigma_i \psi     \label{PCs} \\
    Q_{zi}   &= \frac{\hbar}{2m\im} \Big{(} \psi^\dagger \sigma_i (\partial_z \psi) 
        - (\partial_z \psi^\dagger) \sigma_i \psi \Big{)}, \label{PCQ}
\end{align}
where $\sigma_i$ are the Pauli matrices corresponding to directions $i\in[x',y',z']$ in spin space, where as before $z'~||~\bvec{B}$. In this notation, $Q_{zz'}$ describes the spin current flowing out-of-plane ($z$) with spin direction aligned with $\bvec{B}$ (i.e. $z'$), while $Q_{zx'}$ and $Q_{zy'}$ describe the spin currents flowing out-of-plane with spin direction transverse to $\bvec{B}$. Using \Eqs{WF1}{WF2} and \Eqs{PCs}{PCQ}, the spin density and spin currents near the impurity are:
\begin{align}
    s^0_{i}    &= \chi^\dagger_\inc \big{(} t^\dagger \sigma_i t \big{)} \chi_\inc \label{s0} \\
    Q^{0^-}_{zi}   &= \frac{\hbar k_z}{m} \chi_\inc^\dagger \big{(} \sigma_i - r^\dagger \sigma_i r \big{)} \chi_\inc \label{Q0m} \\
    Q^{0^+}_{zi}   &= \frac{\hbar k_z}{m} \chi^\dagger_\inc \big{(} t^\dagger \sigma_i t \big{)} \chi_\inc \label{Q0p}
\end{align}
The reflection ($r$) and transmission ($t$) matrices are diagonal in spin space,
\begin{align}
    r &= \begin{pmatrix} r_\UA & 0 \\ 0 & r_\DA  \end{pmatrix}, ~~
    t = \begin{pmatrix} t_\UA & 0 \\ 0 & t_\DA  \end{pmatrix},
    \label{rtMat}
\end{align}
where  as before the $\UD$ labels denotes the spin aligned or opposite to the interfacial magnetic field $\bvec{B}$.

\begin{figure}
	\centering
	\vspace{0pt}	
	\includegraphics[width=1\linewidth,trim={0.0cm 0.0cm 0.0cm 0.0cm},clip]{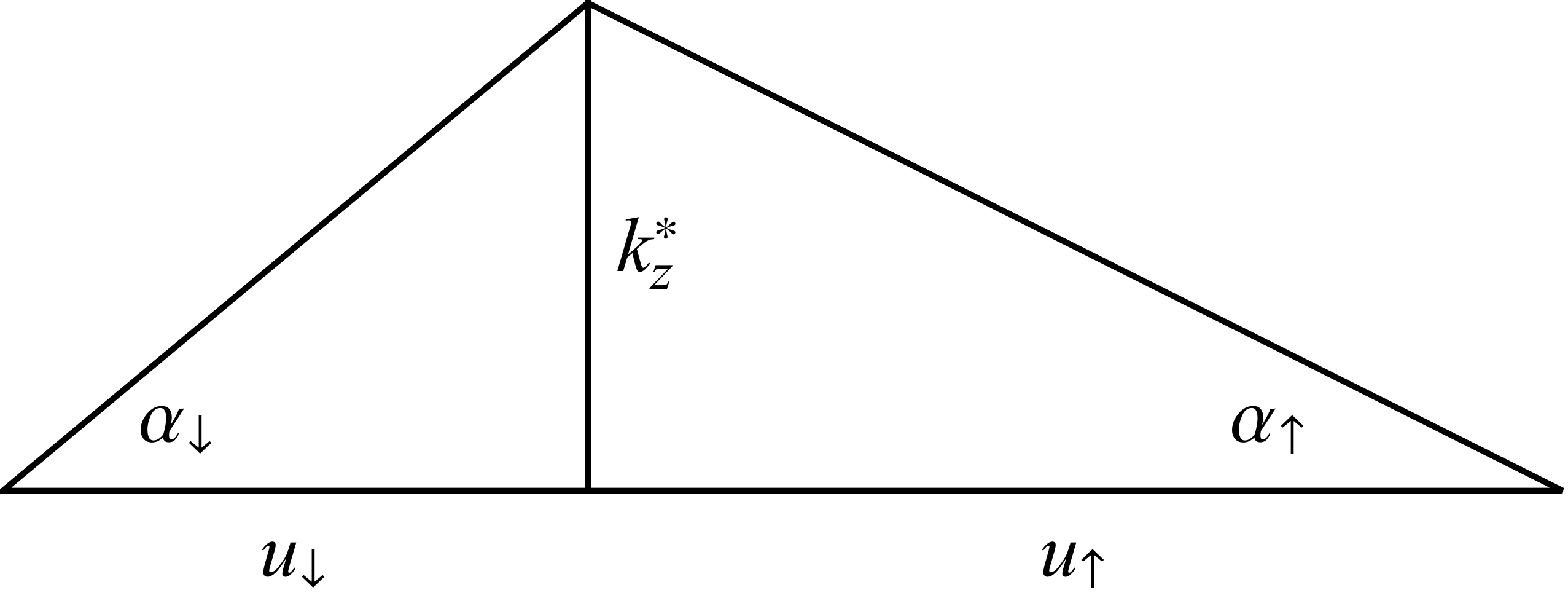}
	\caption{
	Visual representation of the relationship between the angles $\ang_{\UD}$, the barrier strengths $\uUD$, and the scaled $z$-velocity $k_z^* = k_z/k_F$. In the limit that $\uD = 0$ and $\uU \rightarrow \infty$, $\ang_\DA = \pi/2$ and $\ang_\UA \rightarrow 0$.
	}
	\vspace{0pt}
	\label{angles}
\end{figure}

Based on the Hamiltonian given by \Eq{Hmat}, the spin-dependent reflection and transmission amplitudes are:
\begin{align}
    r_{\UD} = \frac{\uUD}{\im \kzB - \uUD} ~~~~
    t_{\UD} = \frac{\im \kzB}{\im \kzB - \uUD}
    \label{rt}
\end{align}
where $\kzB = k_z/k_\text{F}$ is the out-of-plane component of the incident crystal momentum (along $\zhat$) scaled by the Fermi momentum. We can further simplify this notation by introducing an angle $\ang_{\UD}$ (defined geometrically in Fig. \ref{angles}) such that:
\begin{align}
    \cos(\ang_{\UD})    = \frac{\uUD}{\sqrt{(\kzB)^2 + (\uUD)^2}}, \\
    \sin(\ang_{\UD})    = \frac{\kzB}{\sqrt{(\kzB)^2 + (\uUD)^2}}.
    \label{rt2}
\end{align}
Without loss of generality, we may assume that $\kzB$ and $\uUD$ are either zero or positive definite, so that $\ang_{\UD} \in [0,\pi/2]$. The scattering amplitudes then become:
\begin{align}
    r_{\UD} = -e^{\im \ang_{\UD}}\cos(\ang_{\UD}) \label{ra} \\
    t_{\UD} = -\im e^{\im \ang_{\UD}} \sin(\ang_{\UD}) \label{ta}
\end{align}

Using these scattering amplitudes, we may determine the fate of an incident electron spin oriented transverse to the effective magnetic field at the interface. Say the incident electron spin points along $x'$, which corresponds to:
\begin{align}
    \chi_\inc = \frac{1}{\sqrt{2}} \begin{pmatrix} 1 \\ 1 \end{pmatrix}.
    \label{chiinc}
\end{align}
The reflected and transmitted spinors are then:
\begin{align}
    \chi_\refl &= -\frac{1}{\sqrt{2}} 
    \begin{pmatrix} e^{\im \ang_{\UA}}\cos(\ang_{\UA}) \\ e^{\im \ang_{\DA}}\cos(\ang_{\DA}) \end{pmatrix}, \\
    \chi_\trans &= -\frac{\im}{\sqrt{2}} 
    \begin{pmatrix} e^{\im \ang_{\UA}} \sin(\ang_{\UA}) \\ e^{\im \ang_{\DA}} \sin(\ang_{\DA}) \end{pmatrix}.
    \label{chirt}
\end{align}
Let us pause to connect back to the main text, in which $\uU \rightarrow \infty$ and $\uD = 0$. In this limit, $\ang_\UA \rightarrow 0$ and $\ang_\DA = \pi/2$, which gives
\begin{align}
    \chi_\refl &\rightarrow 
    \begin{pmatrix} -1 - \im \ang_\UA \\ 0 \end{pmatrix}
    =
    \begin{pmatrix} -1 - \im \kzB/\uU \\ 0 \end{pmatrix}
    , \\
    \chi_\trans &\rightarrow 
    \begin{pmatrix} -\im \ang_{\UA} \\ 1 \end{pmatrix}
    =
    \begin{pmatrix} -\im \kzB/\uU \\ 1 \end{pmatrix},
    \label{chirt2}
\end{align}
to first order in $\ang_\UA$. Note that we have dropped the normalization constant here. In the previous section, the imaginary part of $\chi_\trans$ gives rise to the transverse spin density required to have a spin torque. Here, we show quantitatively that the absorbed spin current equals the spin torque. 

To verify that the absorbed spin current, given by the discontinuity in spin current across the interface, 
\begin{align}
    \Delta \bvec{Q}_z = \bvec{Q}^{0^-}_{z} -  \bvec{Q}^{0^+}_{z}
\end{align}
equals the spin torque
\begin{align}
    \bvec{\tau} &= 
    \frac{\hbar k_F}{m} (\uU - \uD) \bvec{s}^0 \times \hat{\bvec{B}},
    \label{COAM}
\end{align}
we evaluate the expression $\bvec{\tau} = \Delta \bvec{Q}_z$ using \Eqsd{s0}{Q0p} and \Eqs{ra}{ta}. To prove these two quantities are equal, it is easier to divide both by $k^*_z$, yielding:
\begin{align}
    \bvec{\tau}/k^*_z   &= \Delta \bvec{Q}_z / k^*_z
                        = \frac{\hbar k_F}{m} \begin{pmatrix}  
                        \sin^2(\Delta\ang) \\ 
                        \sin(\Delta\ang) \cos(\Delta\ang) \\ 
                        0 
                        \end{pmatrix} \label{FinalTorque}
\end{align}
The final expression for the torque (scaled by $k^*_z$) depends only on the difference in angles $\Delta \alpha$ and the Fermi momentum. 

From \Eq{FinalTorque} we see that the spin torque at the interface equals the drop in spin current across the interface. The lost spin current was absorbed by the magnetic part of the interface, which resulted in the torque. Furthermore, we see that the spin current component $Q_{zz'}$ is continuous across the interface (i.e. $\Delta Q_{zz'} = 0$).




\section{Phenomenological Theory of Spin Transport at Interfaces with Spin-Orbit Coupling}
\label{appendix:spintransport}

The presence of spin-orbit coupling at interfaces greatly complicates spin transport because spin-orbit coupling opens a channel for angular momentum transfer to and from the atomic lattice. Since nothing in principle restricts the direction of angular momentum flow between conduction electrons and the atomic lattice, interfacial spin-orbit coupling has two consequences: 1) spin currents may give some angular momentum to the atomic lattice when flowing across the interface and 2) the atomic lattice may generate spin currents at the interface. The former is called \emph{spin memory loss} and the latter is called \emph{interface-generated spin currents}.

Our goal is to develop a simple-enough model that qualitatively describes spin memory loss and interface-generated spin currents, as well as other features of spin transport at interfaces with spin-orbit coupling. While quantitative estimates of these phenomena have been obtained from first principles calculations, a simple model helps to introduce the wide variety of phenomena driven and/or influenced by interfacial spin-orbit coupling.

By assuming various boundary conditions, we can qualitatively describe the spin currents and spin torques resulting from both in-plane and out-of-plane electric fields, as well as from spin currents generated elsewhere in the system. Here, boundary conditions refer to our choice of the spin and occupation probability of carriers incident to the interface. Such freedom in boundary conditions enables a description of several important phenomena within the same model, including spin memory loss, interface-generated spin currents, the effect of spin-orbit coupling on the spin mixing conductance, spin transfer torques, and spin-orbit torques.

First, how do we describe the occupation of carriers in a given state? Here, we use a semiclassical description based on the spin-dependent Boltzmann equation. The simplest relevant description, introduced by Camley and Barnas,\cite{CamleyBarnas} assumes carrier spins are parallel or antiparallel to a given axis, and that carriers of each spin species are described by a separate occupation function $f_{\UD}(\bvec{r},\bvec{k})$. The occupation function $f_{\UD}(\bvec{r},\bvec{k})$ is the probability to find a carrier with spin $\uparrow$ or $\downarrow$ at position $\bvec{r}$ with momentum $\bvec{k}$. However, to describe spins along multiple (non-collinear) axes, a more general formalism is required. We could, for instance, assign an occupation function to parallel and antiparallel spins along all three Cartesian axes, giving six occupation functions $f_{is}(\bvec{r},\bvec{k})$ for $i\in[x,y,z]$ and $s \in[\uparrow,\downarrow]$. However, two of these occupation functions are redundant if we only wish to keep track of the spin polarization and the total charge density, in which case we may write four occupation functions instead:
\begin{align}
    f_i(\bvec{r},\bvec{k}) &= f_{i\uparrow} - f_{i\downarrow} \quad \text{for}~i\in[x,y,z]   \\
    f_c(\bvec{r},\bvec{k}) &= \sum_i (f_{i\uparrow} + f_{i\downarrow}).
\end{align}
For simplicity, let us assume that the Boltzmann distribution varies along $z$ but is isotropic along $x$ and $y$. For systems just out of equilibrium, we describe the perturbation of the distribution function as follows
\begin{align}
    f_\alpha(z,\bvec{k}) &= f_\text{eq}(\epsilon_{\bvec{k}}) \delta_{\alpha c} + \frac{\partial f_\text{eq}}{\partial \epsilon_{\bvec{k}}} g_\alpha(z,\bvec{k})
\end{align}
where $\epsilon_{\bvec{k}}$ is the $\bvec{k}$-dependent energy, $f_\text{eq}$ is the (spin-independent) equilibrium distribution function, and $g_\alpha$ is the nonequilibrium perturbation of the distribution function. Note that the distribution functions can be arranged as four-vectors (i.e. $f_\alpha \rightarrow \fourvec{f}$) with components denoted by $\alpha\in[x,y,z,c]$. In the four-vector notation, we have
\begin{align}
    \fourvec{f}(z,\bvec{k}) &= \fourvec{f}_\text{eq}(\epsilon_{\bvec{k}}) + \frac{\partial f_\text{eq}}{\partial \epsilon_{\bvec{k}}} \fourvec{g}(z,\bvec{k})
\end{align}
where
\begin{align}
\fourvec{f}_{\text{eq}}(\epsilon_{\bvec{k}}) &=
\begin{pmatrix}
0		\\
0		\\
0		\\
f_\text{eq}(\epsilon_{\bvec{k}})	
\end{pmatrix},
\quad
\fourvec{g}(z,\bvec{k}) =
\begin{pmatrix}
g_x(z,\bvec{k})		\\
g_y(z,\bvec{k})		\\
g_z(z,\bvec{k})		\\
g_c(z,\bvec{k})		\\
\end{pmatrix}.
\end{align}
The Boltzmann equation is an integro-differential equation that can be used to solve for $\fourvec{g}$ as a function of position and momentum. We omit details of solving the Boltzmann equation here, and instead refer the reader to Ref.~\onlinecite{Xiao:2007}. However, it is important to note that, when solving the Boltzmann equation, boundary conditions are needed at interfaces and these can be supplied by quantum mechanical scattering amplitudes. For instance, at an interface ($z = 0$), the nonequilibrium distribution of incident states is related to the reflected and transmitted states like so
\begin{align}
    \fourvec{g}(0^-,k_x,k_y,-k_z) &= R(\bvec{k}) \fourvec{g}(0^-,k_x,k_y,k_z)   \nonumber \\
    &+ T(\bvec{k}) \fourvec{g}(0^+,k_x,k_y,-k_z)
    \label{ScatDistZm} \\
    \fourvec{g}(0^+,k_x,k_y,k_z) &= T(\bvec{k}) \fourvec{g}(0^-,k_x,k_y,k_z),   \nonumber \\
    &+ R(\bvec{k}) \fourvec{g}(0^+,k_x,k_y,-k_z) 
    \label{ScatDistZp}
\end{align}
where $R(\bvec{k})$ and $T(\bvec{k})$ are $4\times4$ matrices describing reflection and transmission respectively. The $R$ and $T$ matrices used in \Eqs{ScatDistZm}{ScatDistZp} are the same regardless of what layer the carriers are incident from because we assume the layers are identical in equilibrium. We remind the reader that in this model, the nonequilibrium distribution function $\fourvec{g}$ captures the differences in each layer.

We can simplify this notation for spherical Fermi surfaces, where the incident, reflected, and transmitted distribution functions are defined on hemispheres specified by the sign of $k_z$. Thus, we may write 
\begin{align}
    \fourvec{g}^R(0^-,\kp) &= R(\kp) \fourvec{g}^I(0^-,\kp) \label{RmatZm2}   \\
    \fourvec{g}^T(0^+,\kp) &= T(\kp) \fourvec{g}^I(0^-,\kp) \label{TmatZm2}   \\
    \fourvec{g}^R(0^+,\kp) &= R(\kp) \fourvec{g}^I(0^+,\kp) \label{RmatZp2}   \\
    \fourvec{g}^T(0^-,\kp) &= T(\kp) \fourvec{g}^I(0^+,\kp) \label{TmatZp2},
\end{align}
where $\kp = (k_x,k_y)$ is the in-plane crystal momentum of the incoming electrons and the superscripts $I$, $R$ and $T$ denote the incident, reflected and transmitted distribution functions respectively.

The last step in setting up the calculation is to relate the $4\times4$ Boltzmann interface scattering matrices $R$ and $T$ to the $2\times2$ quantum mechanical scattering matrices $r(\kp)$ and $t(\kp)$ that were derived in earlier sections:
\begin{align}
    [R(\kp)]_{\alpha\beta} &= \frac{1}{2} \text{tr}[r^\dagger(\kp) \sigma_\alpha r(\kp) \sigma_\beta]
    \label{R4x4MatApp} \\
    [T(\kp)]_{\alpha\beta} &= \frac{1}{2} \text{tr}[t^\dagger(\kp) \sigma_\alpha t(\kp) \sigma_\beta]
    \label{T4x4MatApp}
\end{align}
We omit the derivation of these equations here, which can be found in Ref.~\onlinecite{iSOCAminPhenomenology}. The expression for the charge and spin currents flowing in direction $i$ ($i \in [x,y,z]$) are
\begin{align}
    j_{i}(z) = \frac{e}{\hbar (2\pi)^3} \int_\text{FS} d\kvec \frac{k_i}{k_F} g_c(z,\bvec{k}) \\
    Q_{is}(z) = \frac{1}{2 (2\pi)^3} \int_\text{FS} d\kvec \frac{k_i}{k_F} g_s(z,\bvec{k})
\end{align}
in units of ${\rm A/m^2}$ (charge current density) and ${\rm J/m^2}$ (angular momentum current density) respectively. We can combine these definitions into a single definition
\begin{align}
    j_{i\alpha}(z) = c_j \int_\text{FS} d\kvec \frac{k_i}{k_F} g_\alpha(z,\bvec{k})
\end{align}
where $c_j = e/\hbar (2\pi)^3$ and $\alpha \in [x,y,z,c]$ as before. Note that the spin current tensor elements ($\alpha = x,y,z$ for any $i$) are given in units of charge current density and can be converted back to an angular momentum current density by multiplying by $\hbar/2e$. In the main text, we also define the spin/charge accumulation at $z = 0$ using the constant $c_\mu = -1/4\pi e k_F^2$.

In what follows, we are only interested in the out-of-plane flowing charge and spin currents (i.e.~along $\zhat$). We can then rewrite the above expression in four-vector notation at the interface as:
\begin{align}
    \fourvec{j}_{z}(0^\pm) &= c_j \int_\text{FS} d\kvec \frac{k_z}{k_F}\fourvec{g}(0^\pm,\kvec).
\end{align}
We know that the incident distribution functions (defined on one hemisphere of the Fermi surface) at $z = 0^\pm$ are related to the reflected and transmitted distribution functions (defined on the other hemisphere) by \Eqsd{RmatZm2}{TmatZp2}. Thus we may write the total spin/charge currents in terms of the incident, reflected, and transmitted contributions as follows,
\begin{align}
    \fourvec{j}_{z}(0^\pm) &= \mp \big{(}
    \fourvec{j}^\text{I}_{z}(0^\pm) - 
    \fourvec{j}_{z}^\text{R}(0^\pm) - 
    \fourvec{j}_{z}^\text{T}(0^\pm) \big{)} \\
    &= \mp c_j \int_\text{2DBZ} d\kp 
    ( I - R ) \fourvec{g}^\text{I}(0^\pm)
    - T \fourvec{g}^\text{I}(0^\mp)
    \label{jzfinal}
\end{align}
where the last line is rewritten as an integral over the two-dimensional Brillouin zone (2DBZ) spanned by $k_x$ and $k_y$. Note that the $\kp$-dependence of $R$, $T$, and $\fourvec{g}^\text{I}$ has been omitted for simplicity.

As before, we assume that carriers see an effective magnetic field $\bvec{B}(\bvec{k}) = \bvec{B}_\text{ex} + \bvec{B}_\text{soc}(\bvec{k})$ at the interface, where $\bvec{B}_\text{ex}$ is the exchange field and $\bvec{B}_\text{soc}(\bvec{k})$ is the momentum-dependent spin-orbit field. A free electron gas describes the bulk of each layer while a delta function potential describes the interface. If $\bvec{B}(\bvec{k})$ points along $\zhat'$ (which corresponds to the spin reference frame), then the $R$ and $T$ matrices computed using \Eqs{R4x4MatApp}{T4x4MatApp} are given by:
\begin{align}
\bar{R} &=
\begin{pmatrix}
1-a_\irl	&	b_\irl  	&	0	    &	0	\\
-b_\irl		&	1-a_\irl	&	0	    &	0	\\
0   		&	0	    	&	1-c	    &	-d	\\
0   		&	0	    	&	-d	    &	1-c	
\end{pmatrix},
\\
\bar{T} &=
\begin{pmatrix}t
a_\tl		&	-b_\tl  	&	0	    &	0	\\
b_\tl		&	a_\tl		&	0	    &	0	\\
0   		&	0	    	&	c	    &	d	\\
0   		&	0	    	&	d	    &	c	
\end{pmatrix}
\end{align}
where all tensor elements are real-valued and depend on $k_x$ and $k_y$. The parameters $c$ and $d$ are identical in both tensors and are a consequence of particle conservation during scattering. Note that the matrices $I - R$ and $T$ both have the following form
\begin{align}
\begin{pmatrix} 
a_\irtl		&	-b_\irtl  	&	0	    &	0	\\
b_\irtl		&	a_\irtl		&	0	    &	0	\\
0   		&	0	    	&	c	    &	d	\\
0   		&	0	    	&	d	    &	c	
\end{pmatrix},
\end{align}
where $\irtl \in [\irl,\tl]$. If $\bvec{B}(\bvec{k})$ points along some general direction, the scattering matrices become
\begin{align}
R(\bvec{k}) = O(\bvec{k}) \bar{R}(\bvec{k}) O(\bvec{k})^\dagger  \label{Rscat} \\
T(\bvec{k}) = O(\bvec{k}) \bar{T}(\bvec{k}) O(\bvec{k})^\dagger
\label{Tscat}
\end{align}
where $O$ is any orthogonal transformation rotating the vector $\zhat$ to the direction parallel to $\bvec{B}(\bvec{k})$. By switching to spherical coordinates
\begin{align}
k_x &= k_F \sin(\theta)\cos(\phi) \\
k_y &= k_F \sin(\theta)\sin(\phi) \\
k_z &= k_F \cos(\theta),
\end{align}
it becomes apparent that the $\bar{R}$ and $\bar{T}$ matrices only depend on $\theta$ while the orthogonal transformations $O$ encode the $\phi$-dependence. This is because the $2\times2$ reflection and transmission matrices defined in \Eq{rt} that are used to calculate $\bar{R}$ and $\bar{T}$ depend only on $k_z$, or alternatively, only on $\theta$. Thus, we may write: 
\begin{align}
R(\theta,\phi) = O(\theta,\phi) \bar{R}(\theta) O(\theta,\phi)^\dagger  \label{Rscat2} \\
T(\theta,\phi) = O(\theta,\phi) \bar{T}(\theta) O(\theta,\phi)^\dagger.
\label{Tscat2}
\end{align}
In spherical coordinates we can more easily write the explict form of the $O$ matrices. For an out-of-plane magnetization, these matrices are given by
\begin{align}
O(r,\phi) &= 
\begin{pmatrix}
1	&	0       	    	&	0	                &	0	\\
0   &   \fs(\theta,\uangle)		&	-\fc(\theta,\uangle)      &	0	\\
0   &   \fc(\theta,\uangle)		&	\fs(\theta,\uangle)       &	0	\\
0	&	0       	    	&	0	                &	1	
\end{pmatrix} \nonumber \\
&\quad \times
\begin{pmatrix}
\cos(\phi)		&	\sin(\phi)	    &	0	    &	0	\\
0	        	&	0	        	&	1  	    &	0	\\
\sin(\phi)		&	-\cos(\phi)		&	0	    &	0	\\
0	           	&	0       		&	0	    &	1	
\end{pmatrix},
\end{align}
where the functions $\fsf$ and $\fcf$ are given in \Eq{SCI}.
We remind the reader that $\uangle$ encodes the relative dependence on the interfacial exchange and spin-orbit interactions, where $u_\text{ex} = |u_\text{eff}|\cos(\uangle)$ and $u_\text{R} = |u_\text{eff}|\sin(\uangle)$.

Rewriting $\fourvec{j}_z(0^\pm)$ in spherical coordinates gives
\begin{align}
    \fourvec{j}_{z}(0^\pm) = 
    \mp c_j k_F^2 \int &d\theta \sin(\theta)
    \int d\phi \\
    &\times
    ( I - R ) \fourvec{g}^\text{I}(0^\pm)
    - T \fourvec{g}^\text{I}(0^\mp).
    \label{jzSpherical}
\end{align}
As seen in the main text, it is convenient to analyze the spin/charge currents and the distribution functions in terms of their average values and difference in values across the interface, defined in \Eq{DiffAvgdef}.
Some algebra gives:
\begin{align}
    \Delta\fourvec{j}_{z} = 
    c_j k_F^2 \int &d\theta \sin(\theta)
    \int d\phi \\
    &\times
    ( I - R - T) \bar{\fourvec{g}}^\text{I} \\
    \bar{\fourvec{j}}_{z} = 
    c_j k_F^2 \int &d\theta \sin(\theta)
    \int d\phi \\
    &\times
    ( I - R + T) \Delta\fourvec{g}^\text{I}
    \label{jzSphericalMP}
\end{align}
Performing the $\phi$ integral is tedious but straightforward, while performing the $\theta$ integral is much more difficult and does not change the conceptual understanding of the model. In this spirit, we define the following matrices:
\begin{align}
    \SMir^n &= \int d\phi \cos^n(\phi) \big{(} I - R(\theta,\phi) \big{)} \\
            &= \int d\phi \cos^n(\phi) \big{(} I - O(\theta,\phi) \bar{R}(\theta) O(\theta,\phi)^\dagger \big{)} \\
    \SMt^n  &= \int d\phi \cos^n(\phi) T(\theta,\phi) \\
            &= \int d\phi \cos^n(\phi) O(\theta,\phi) \bar{T}(\theta) O(\theta,\phi)^\dagger
\end{align}
The result of evaluating these integrals yields the expressions in \Eq{Aforz} (for $n = 0$) and \Eq{Aforx} (for $n = 1$) in the main text. Using the definition $v(\theta) = c_j e k_F^2 \sin(\theta)$, we may then write:
\begin{align}
    \Delta\fourvec{j}_{z} &= 
    \int d\theta v(\theta) (\SMir^0 - \SMt^0) \bar{\fourvec{\pu}} \\
    \bar{\fourvec{j}}_{z} &= 
    \int d\theta v(\theta) (\SMir^0 + \SMt^0) \Delta\fourvec{\pu}
    \label{jzSphericalforZ}
\end{align}
when $\fourvec{g}(0^\pm,\bvec{k}) = e \fourvec{\pu}_\pm$ as defined in \Eq{gacc} and
\begin{align}
    \Delta\fourvec{j}_{z} &= 
    \int d\theta v(\theta) (\SMir^1 - \SMt^1) \bar{\fourvec{\pu}} \\
    \bar{\fourvec{j}}_{z} &= 
    \int d\theta v(\theta) (\SMir^1 + \SMt^1) \Delta\fourvec{\pu}
    \label{jzSphericalforX}
\end{align}
when $\fourvec{g}(0^\pm,\bvec{k}) = e \tilde{k}_x \fourvec{\pu}_\pm = e \cos(\phi) \fourvec{\pu}_\pm$ as defined in \Eq{gip}. The forms of both \Eqs{jzSphericalforZ}{jzSphericalforX} are quite similar, so following the main text, we write
\begin{align}
    \Delta\fourvec{j}_{z} &= 
    \int d\theta v(\theta) \Delta\SM(\theta) \bar{\fourvec{\pu}} \\
    \bar{\fourvec{j}}_{z} &= 
    \int d\theta v(\theta) \bar{S}(\theta) \Delta\fourvec{\pu}
    \label{jzSphericalforZX}
\end{align}
for both choices of $\fourvec{g}$, where the symmetric and antisymmetric matrices are defined as
\begin{align}
    \Delta\SM = \SMir - \SMt \\
    \bar{\SM} = \SMir + \SMt
\end{align}
and the explicit form of $\SMir$ and $\SMt$ depends on the choice of $\fourvec{g}$ as seen above.


\providecommand{\noopsort}[1]{}\providecommand{\singleletter}[1]{#1}%

\end{document}